\documentclass[lettersize,journal]{IEEEtran}
\usepackage{amsmath,amssymb,amsfonts}
\usepackage{algorithmic}
\usepackage{algorithm}
\usepackage{array}
\usepackage[caption=false,font=normalsize,labelfont=sf,textfont=sf]{subfig}
\usepackage{textcomp}
\usepackage{stfloats}
\usepackage{url}
\usepackage{verbatim}
\usepackage{graphicx}
\usepackage{cite}
\usepackage{booktabs}
\usepackage{xcolor}
\usepackage{epsfig}
\usepackage{epstopdf}
\usepackage{acronym}
\graphicspath{{fig/}}
\usepackage{mathtools}
\usepackage{amsthm}

\usepackage{ragged2e}

\newacro{ACDD}{Alamouti with cyclic delay diversity}
\newacro{URLLC}{ultra-reliable low-latency communications}
\newacro{3GPP}{third generation partnership project}
\newacro{PHY}{physical layer}
\newacro{MIMO}{multiple-input multiple-output}
\newacro{SIMO}{single-input multiple-output}
\newacro{MISO}{multiple-input single-output}
\newacro{SISO}{single-input single-output}
\newacro{MRC}{maximum-ratio combining}
\newacro{SNR}{signal-to-noise ratio}
\newacro{CP}{cyclic prefix}
\newacro{CDD}{cyclic delay diversity}
\newacro{FSC}{frequency-selective channel}
\newacro{STC}{space-time coding}
\newacro{FFT}{fast Fourier transform}
\newacro{LMMSE}{linear minimum mean-squared error}
\newacro{FER}{frame error rate}
\newacro{OFDM}{orthogonal frequency division multiplexing}
\newacro{OCDM}{orthogonal chirp division multiplexing}
\newacro{FSC}{frequency-selective channel}
\newacro{CSI}{channel state information}
\newacro{LMMSE-PIC}{linear minimum mean squared error with parallel interference cancellation}
\newacro{PFE}{perfect-feedback equalizer}
\newacro{FD}{frequency domain}
\newacro{PDP}{power delay profile}
\newacro{PDF}{probability density function}
\newacro{DFT}{discrete Fourier transform}
\newacro{SDFT}{sparse DFT}
\newacro{ICI}{inter-carrier interference}
\newacro{OTFS}{orthogonal time frequency space}
\newacro{AWGN}{additive white Gaussian noise}
\newacro{SWH}{sparse Walsh-Hadamard}
\newacro{LLR}{log-likelihood ratio}
\newacro{PMF}{probability mass function}
\newacro{CRC}{cyclic redundancy check}
\newacro{PAM}{pulse amplitude modulation}
\newacro{QAM}{quadrature amplitude modulation}
\newacro{FWHT}{fast Walsh-Hadamard transform}
\newacro{MAP}{maximum a-posteriori}
\newacro{SC}{single-carrier}
\newacro{ISI}{inter-symbol interference}
\newacro{ZP}{zero-padding}
\newacro{BCJR}{Bahl, Cocke, Jelinek, and Raviv}
\newacro{WHT}{Walsh-Hadamard transform}
\newacro{APP}{a-posteriori probability}
\newacro{SILE-EPIC}{self-iterated linear equalizer with expectation propagation}
\newacro{EP}{expectation propagation}
\newacro{i.i.d.}{independent and identically distributed}
\newacro{CWCU}{component wise conditionally unbiased}
\newacro{MSE}{mean squared error}
\newacro{EXIT}{extrinsic information transfer}
\newacro{MI}{mutual information}
\newacro{PAPR}{peak-to-average power ratio}
\newacro{DFT-s}{discrete Fourier transform-spread}
\newacro{AMP}{approximate message passing}
\newacro{GAMP}{generalized \ac{AMP}}
\newacro{VAMP}{vector \ac{AMP}}
\newacro{RSC}{recursive systematic convolutional}
\newacro{QPSK}{quadrature phase-shift keying}

\newcommand{\ma}[1]{\mathbf{#1}}

\newcommand{\pa}[1]{\left(#1\right)}

\def\BibTeX{{\rm B\kern-.05em{\sc i\kern-.025em b}\kern-.08em
		T\kern-.1667em\lower.7ex\hbox{E}\kern-.125emX}}
	
\usepackage{soul,color}

\usepackage{tikz}
\usetikzlibrary{shapes,arrows}
\usetikzlibrary{positioning,calc}
\usetikzlibrary{decorations.pathreplacing,calligraphy}
\usetikzlibrary{arrows.meta}

\tikzset{add/.style n args={4}{
		minimum width=3mm,
		path picture={
			\draw[black] 
			(path picture bounding box.south east) -- (path picture bounding box.north west)
			(path picture bounding box.south west) -- (path picture bounding box.north east);
			\node at ($(path picture bounding box.south)+(0,0.13)$)     {\tiny #1};
			\node at ($(path picture bounding box.west)+(0.13,0)$)      {\tiny #2};
			\node at ($(path picture bounding box.north)+(0,-0.13)$)        {\tiny #3};
			\node at ($(path picture bounding box.east)+(-0.13,0)$)     {\tiny #4};
		}
	}
}

\tikzset{add2/.style n args={4}{
		minimum width=1mm,
		path picture={
			\draw[black] 
			(path picture bounding box.south) -- (path picture bounding box.north)
			(path picture bounding box.west) -- (path picture bounding box.east);
			\node at ($(path picture bounding box.south)+(0,0.13)$)     {\tiny #1};
			\node at ($(path picture bounding box.west)+(0.13,0)$)      {\tiny #2};
			\node at ($(path picture bounding box.north)+(0,-0.13)$)        {\tiny #3};
			\node at ($(path picture bounding box.east)+(-0.13,0)$)     {\tiny #4};
		}
	}
}

\newcounter{MYtempeqncnt}

\begin{document}

\title{Sparse-DFT and WHT Precoding with Iterative Detection for Highly Frequency-Selective Channels
 \thanks{Roberto Bomfin is with the Engineering Division, New York University
(NYU) Abu Dhabi, 129188, UAE (email: roberto.bomfin@nyu.edu).}
\thanks{Marwa Chafii is with Engineering Division, New York University (NYU)
Abu Dhabi, 129188, UAE and NYU WIRELESS, NYU Tandon School of
Engineering, Brooklyn, 11201, NY, USA (email: marwa.chafii@nyu.edu).}}

\author{Roberto Bomfin and Marwa Chafii}



\maketitle
	
\begin{abstract}
Various precoders have been recently studied by the wireless community to combat the channel fading effects.
Two prominent precoders are implemented with the discrete Fourier transform (DFT) and Walsh-Hadamard transform (WHT).
The WHT precoder is implemented with less complexity since it does not need complex multiplications.
Also, spreading can be applied sparsely to decrease the transceiver complexity, leading to sparse DFT (SDFT) and sparse Walsh-Hadamard (SWH).
Another relevant topic is the design of iterative receivers that deal with inter-symbol-interference (ISI).
In particular, many detectors based on expectation propagation (EP) have been proposed recently for channels with high levels of ISI.
An alternative is the maximum a-posterior (MAP) detector, although it leads to unfeasible high complexity in many cases.
In this paper, we provide a relatively low-complexity \textcolor{black}{computation} of the MAP detector for the SWH. We also propose two \textcolor{black}{feasible methods} based on the Log-MAP and Max-Log-MAP.
Additionally, the DFT, SDFT and SWH precoders are compared using an EP-based receiver with one-tap FD equalization.
Lastly, SWH-Max-Log-MAP is compared to the (S)DFT with EP-based receiver in terms of performance and complexity.
The results show that the proposed SWH-Max-Log-MAP has a better performance and complexity trade-off for QPSK and 16-QAM under highly selective channels, but has unfeasible complexity for higher QAM orders.
\end{abstract}

\begin{IEEEkeywords}
Single Carrier, Walsh Hadamard, DFT-spread-OFDM, maximum a-posteriori estimation, expectation propagation
\end{IEEEkeywords}

\section{Introduction}
\IEEEPARstart{S}{everal} waveforms have been investigated in the past years for wireless communications.
The most common one employed in modern wireless systems is \ac{OFDM} due to its simple frequency-domain (FD) equalization and scalability.
However, alternative approaches have been investigated to mitigate the effect of time and/or frequency selective channels \cite{OTFS,BomfinTWC,OCDM,Zemen,ViterboWHT,BomfinSWH,Marwa}.
For instance, a common alternative to OFDM is to perform \ac{DFT} precoding in the frequency domain (FD) symbols.
This procedure is called DFT-spread-OFDM and has been used in the uplink of 4G and 5G due to its low \ac{PAPR} \cite{PAPR2}.
In this work, we refer to DFT-spread-OFDM as DFT precoding.
In addition to the advantageous PAPR aspect of DFT precoding, it effectively allows a higher data rate transmission than OFDM for the same SNR.
The authors in \cite{Carmon} have shown that \ac{SC}, which can be seen as DFT-spread-OFDM with zero padding instead of \ac{CP}, offers a higher data rate than OFDM.
Also, since DFT precoding can be seen as a spreading waveform that spreads the symbols energy in the FD, it maximizes the data rate if the receiver can completely remove \ac{ISI} \cite{Bomfin}.
In this paper, we focus on sparse and full FD precoding for highly frequency-selective channels only, \textcolor{black}{which happens in non line-of-sight (NLoS) propagation environments where the delayed replicas of the transmitted signal have significant power contribution and there is no path with dominant power, e.g., indoor factory scenarios with metal structures and machinery \cite{Burmeister}.}
\textcolor{black}{Nevertheless, as shown in \cite{BomfinSWH,BomfinTWC}, the precoder can be designed to account for time spreading as well, where the energy of symbols are spread along several transmission sub-blocks with each block having a \ac{FFT} plus \ac{CP} size.}
Therefore, this work could be extended in this direction.

Besides the waveforms based on the \ac{DFT} precoding such as DFT-spread-OFDM, the \ac{WHT} has been also considered recently as a more attractive precoding alternative from the hardware implementation point of view\cite{WH1,Zemen,ViterboWHT}.
This is because WHT requires no complex multiplications to be implemented as opposed to the DFT, while it does achieve the equal gain criterion.
Other variants employing WHT-based precoding for doubly dispersive channels using a similar concept to OTFS have been also studied \cite{Bomfin,ViterboWHT}.
Another approach to decrease the processing complexity at the receiver side is to employ sparse spreading instead of full spreading.
In particular, the authors in \cite{BomfinSWH} have shown that for $Q>L$, $L$ being the channel length in samples, the spreading can be done sparsely over $Q$ sub-carriers such that all symbols still experience the same channel gain.
\textcolor{black}{The advantage of the sparse spreading is that the receiver now can be implemented with transformations of smaller size, which decreases the complexity regardless of the receiver type.}
In this paper, we also investigate the impact of sparse spreading for DFT precoding, i.e., \ac{SDFT}.

It is known that the spreading waveforms discussed above require iterative equalization in order to take the maximum advantage of the spreading feature \cite{TURBO3,Bomfin_WCNC2,Long,BomfinTWC,OCDM,Zemen,ViterboWHT,BomfinSWH}, otherwise, the performance is degraded significantly. 
This type of receiver was first referred to as turbo equalization in \cite{TURBO_EQ}, where the same idea of turbo decoding is applied to the equalization process at the symbol level.
Although the basic structure of the receiver does not change, there is a vast number of implementation possibilities. 
\textcolor{black}{For instance, MAP detection \cite{TURBO_EQ} computes the exact a-posteriori symbol probability and theoretically represents the best solution in terms of performance}, however, this approach leads to unfeasible complexity in general.
Alternative approaches employ a linear equalizer (LE) based on the \ac{LMMSE-PIC} for the symbol estimation \cite{TURBO_EQ_linear_eq,Max,Studer,Fang,TURBO3,Bomfin_WCNC2}, where the a-priori symbols are used to cancel interference, i.e., coupling among the data symbols. \textcolor{black}{In comparison to the MAP equalizer, the LMMSE-PIC assumes that the data symbols are Gaussian distributed which leads to linear equalization.}
A more general class of detectors considers an additional expectation propagation (EP)-based step, where the EP-based feedback to the LMMSE equalizer refines the soft symbols estimation \cite{Sahin,Sahin2,Santos,Santos2,Dan,Wu}. 
\textcolor{black}{The EP-based receivers with linear equalization also assume that the symbols are Gaussian distributed, but are superior to the LMMSE-PIC due to an additional refinement process to improve the symbol estimates}.
\textcolor{black}{In summary, while the MAP based receivers provide significantly better performance than LE for highly frequency-selective channels \cite{TURBO3}, the EP-based receivers provide a good performance and complexity trade-off between MAP and single LMMSE detection methods \cite{Sahin}}.

Although the MAP equalizer is typically highly complex, we have derived in our previous work \cite{BomfinMAP} a MAP equalizer with decreased complexity for the SWH waveform of \cite{BomfinSWH}. 
\textcolor{black}{Differently from the DFT-based transform, the SWH allows independent in-phase and quadrature processing, and it has been shown in \cite{BomfinMAP} that SWH modulation generates amplitudes that belong to a small set.
The MAP detector can benefit from this fact to avoid unnecessary multiplications to decrease its complexity.}
\textcolor{black}{However, our work in \cite{BomfinMAP} does not discuss the computation of the exponential terms in the symbols marginalization, or \acp{LLR} computation, which can be a bottleneck in terms of complexity since the exponential function requires multiplications to be implemented.
In this paper, two feasible methods based on the Log-MAP and Max-Log-MAP \cite{Robertson} are investigated that avoids computing the exponential function explicitly.}
The Log-MAP method directly computes the extrinsic \acp{LLR} by evaluating the logarithm of exponential summation (LogSumExp) recursively.
This method uses a pre-stored table which avoids the explicit computation of the exponential and logarithm functions.
The Max-Log-MAP solution avoids the look-up table operation and reduces the complexity by approximating the LogSumExp by the maximum argument of the exponential functions.
Although the overall complexity is considerably reduced due to a small number of multiplications, the number of additions to compute the extrinsic \acp{LLR} can still be prohibitively high depending on the system parameters. 
This bottleneck is identified in this paper and a future research direction to mitigate this issue is given.

There have been many works on EP-based receivers. In particular, authors in \cite{Sahin} have proposed a self-iterated \ac{FD} linear equalizer-interference cancellation with EP-based feedback (SILE-EPIC) applied to the DFT precoded system, where the data symbols are assumed to belong to a multivariate white Gaussian distribution.
This assumption allows the one-tap equalization step in FD, {which considerably decreases} the receiver's complexity.
A similar idea has been reported in \cite{Bomfin_WCNC2} for the \ac{LMMSE-PIC}, where this simplification leads to no performance loss.
\textcolor{black}{Most importantly, the SILE-EPIC of \cite{Sahin} provides a better performance and complexity trade-off than other EP-based receivers. 
In particular, the receivers in \cite{Dan,Wu} are sub-optimal due to a single self-iteration (SI) and absence of damping.
The receiver of \cite{Santos2} slightly outperforms the SILE-EPIC, but with much higher computational complexity. 
Moreover, the SILE-EPIC has shown superior performance than the generalized approximate message passing (GAMP) \cite{GAMP}  and vector AMP (VAMP) \cite{VAMP} with extrinsic output.
Thus, in this work, we use the SILE-EPIC of \cite{Sahin} as the FD EP-based benchmark receiver.}
We highlight that other precoders besides DFT have not been evaluated in the original SILE-EPIC paper.
\textcolor{black}{Thus, in this work, we modify the SILE-EPIC to work with the SDFT and SWH precoders, where detailed performance and complexity comparisons among these precoders are provided.
	In particular, we show that the SDFT system has a smaller mean squared error of the estimated symbols than the DFT system, which leads to slightly better performance.}

In this paper, we provide two \textcolor{black}{feasible methods} based on the Log-MAP and Max-Log-MAP for the SWH precoder.
Additionally, the DFT, SDFT, and SWH precoders are compared using SILE-EPIC receivers with one-tap FD equalization.
Lastly, SWH-Max-Log-MAP is compared to the (S)DFT with EP-based receiver in terms of performance and complexity. 

The contributions of this paper are summarized as follows:
\textcolor{black}{\begin{itemize}
	\item Extend our work in \cite{BomfinMAP} regarding the MAP detector for the SWH, by developing and analyzing two feasible methods that avoid the computation of exponential terms in the LLR computation, namely, the Log-MAP and Max-Log-MAP. The results show that the Max-Log-MAP presents a small performance loss and the \textcolor{black}{Log-MAP} performs equally to the exact one of \cite{BomfinMAP}.
	\item \textcolor{black}{Complement the work of \cite{Sahin} by extending the SILE-EPIC receiver to operate with SDFT and WHT precoders. The results show that the sparse precoders decrease the receiver's complexity due to the transformation of smaller size, which is more significant for \ac{QPSK}. In terms of performance, it is shown that the SDFT provides slightly better performance than DFT.}
	\item Provide a complexity analysis of Log-MAP and Max-Log-MAP equalizers for SWH, and SILE-EPIC in terms of multiplications and additions per QAM symbol, such that the different methods can be compared fairly.
	\item Provide a detailed performance vs. complexity trade-off comparison between DFT and WHT-based precoders employing the SILE-EPIC or MAP-based receivers. The results show that SWH-Max-Log-MAP has a better performance and complexity trade-off for QPSK and 16-QAM, while for higher QAM orders it has unfeasible complexity.
\end{itemize}}

The remainder of this paper is organized as follows. 
In Section \ref{sec:system_model}, we provide a general description of the system model in FD.
In Section \ref{sec:sparse_precoders}, the system model for the sparse precoders is given in detail.
In Section \ref{sec:map_detectors_SWH}, the Log-MAP and Max-Log-MAP methods are provided.
In Section \ref{sec:sile_epic}, the SILE-EPIC receiver of \cite{Sahin} is described and the modifications for SDFT and SWH are provided.
In Section \ref{sec:complexity_analysis}, a detailed complexity analysis in terms of real additions and multiplications per \ac{QAM} symbol is given for the SWH-MAP and SILE-EPIC receivers.
In Section \ref{sec:numerical_evaluation}, a numerical evaluation is provided for performance and complexity analysis.
Finally, Section \ref{sec:conclusion} concludes the paper and provide ideas for future research.

{\it Notations:} column vectors are described as lower case bold symbols as $\ma{x}$ whose $n$th element is $x_n$. 
Matrices are described as upper case bold symbols as $\ma{X}$, \textcolor{black}{whose element in the $n$th row and $m$th column is given by $X_{n,m}$.}
The transpose and conjugate transpose are defined respectively as $(\cdot)^{\rm T}$ and $(\cdot)^{\rm H}$.
The \ac{QAM} and \ac{PAM} sets are defined respectively as $\mathcal{D}$ and $\mathcal{R}$.
\textcolor{black}{The set of complex number is defined as $\mathbb{C}$}. The $N \times N$ matrices $\ma{I}_N$, $\ma{F}_N$ and $\ma{W}_N$ represent the identity, normalized Fourier and Walsh-Hadamard matrices, respectively, \textcolor{black}{such that $\ma{F}_N ^{\rm H} \ma{F}_N = \ma{W}_N ^{\rm H} \ma{W}_N = \ma{I}_N$.}
Other functions and symbols are defined as they occasionally appear.

\section{System Model}\label{sec:system_model}
\subsection{Transmitter}\label{subsec:transmitter}
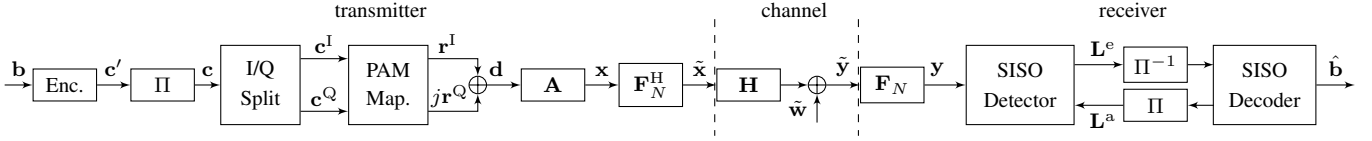
\begin{figure*}[t!]
	\tikzstyle{block} = [draw, fill=white, rectangle, 
minimum height=3em, minimum width=3em]
\tikzstyle{block2} = [draw, fill=white, rectangle, 
minimum height=1em, minimum width=2.4em]

\tikzstyle{multiplier} = [draw,circle,scale=0.7,add={}{}{}{}] {} 
\tikzstyle{sum} = [draw,circle,scale=0.7,add2={}{}{}{}] {} 
\tikzstyle{input} = [coordinate]
\tikzstyle{output} = [coordinate]
\tikzstyle{pinstyle} = [pin edge={to-,thin,black}]

\def\windup{
	\tikz[remember picture,overlay]{
		\draw (-0.8,0) -- (0.8,0);
		\draw (-0.8,-0.4)--(-0.4,-0.4) -- (0.4,0.4) --(0.8,0.4);
}}

\centering
\begin{tikzpicture}[auto, node distance=2cm,>=latex']

	\node [input, name=input] at (0,0) {};
	
	\node [block2, right of=input,node distance=0.8cm] (enc) {\color{black}{\footnotesize Enc.}};
	
	\node [block2, right of=enc,node distance=1.3cm] (interleaver) {\color{black}{\footnotesize $\Pi$}};
	
	\node [block, right of=interleaver,node distance=1.3cm] (split) {\color{white}{\footnotesize Split}};
	\node [ right of=interleaver,node distance=1.3cm,yshift=0.2cm] (split1) {\color{black}{\footnotesize I/Q}};
	\node [ right of=interleaver,node distance=1.3cm,yshift=-0.2cm] (split2) {\color{black}{\footnotesize Split}};
	
	\node [block, right of=split,node distance=1.7cm] (mapper) {\color{white}{\footnotesize Map.}};
	\node [ right of=split,node distance=1.7cm,yshift=0.2cm] (mapper1) {\color{black}{\footnotesize PAM}};
	\node [ right of=split,node distance=1.7cm,yshift=-0.2cm] (mapper2) {\color{black}{\footnotesize Map.}};

	\node [input, right of = mapper,node distance=0.53cm, yshift=0.35cm,name=mapper_auxtop] {};
	\node [input, right of = mapper,node distance=0.53cm, yshift=-0.35cm,name=mapper_auxbottom] {};
	
	\node [ sum,right of=mapper,node distance=1.7cm] (sum) {};
	
	\node [block2, right of=sum,node distance=1.0cm] (modulator) {\color{black}{\footnotesize $\ma{A}$}};
	
	\node [block2, right of=modulator,node distance=1.3cm] (ifft) {\color{black}{\footnotesize $\ma{F}^{\rm H}_N$}};
	
	\node [block2, right of=ifft,node distance=1.3cm] (channel) {\color{black}{\footnotesize $\ma{H}$}};
	
	\node [ sum,right of=channel,node distance=1.3cm] (noise_sum) {};
	\node [ input,below of=noise_sum,node distance=0.5cm] (noise) {};
	
	\node [block2, right of=noise_sum,node distance=1.0cm] (fft) {\color{black}{\footnotesize $\ma{F}_N$}};
	
	\node [output, right of = sum,node distance=0.6cm, name=output] {};

	\draw [->] (input) -- node [xshift=0cm]{\footnotesize $\ma{b}$} (enc);
	\draw [->] (enc) -- node [xshift=0cm]{\footnotesize $\ma{c}'$} (interleaver);
	\draw [->] (interleaver) -- node [xshift=0cm]{\footnotesize $\ma{c}$} (split);
	
	\begin{scope}[transform canvas={yshift=0.35cm}]
		\draw [->] [](split) -- node [yshift=-0.05cm]{\footnotesize$\ma{c}^{\rm I}$}(mapper);
	\end{scope}
	\begin{scope}[transform canvas={yshift=-0.35cm}]
		\draw [->] [](split) -- node [yshift=-0.05cm]{\footnotesize$\ma{c}^{\rm Q}$}(mapper);
	\end{scope}
	
	\draw [->] (mapper_auxtop) -| node [xshift=-0.65cm,yshift=0.2cm]{\footnotesize$\ma{r}^{\rm I}$} (sum);
	\draw [->] (mapper_auxbottom) -| node [xshift= 0cm,yshift=0.2cm]{\footnotesize$j\ma{r}^{\rm Q}$} (sum);	
	
	\draw [->] (sum) -- node [xshift=-0.1cm]{\footnotesize$\ma{d}$} (modulator);
	\draw [->] (modulator) -- node [xshift=.0cm]{\footnotesize${\ma{x}}$} (ifft);
	\draw [->] (ifft) -- node [xshift=.0cm]{\footnotesize${\tilde{\ma{x}}}$} (channel);
	\draw [->] (channel) -- node [xshift=.0cm]{} (noise_sum);
	\draw [->] (noise) -- node [xshift=.0cm]{\footnotesize${\tilde{\ma{w}}}$} (noise_sum);
		\draw [->] (noise_sum) -- node [xshift=.0cm]{\footnotesize${\tilde{\ma{y}}}$} (fft);
	
	
	\node [block, right of=fft,node distance=1.7cm] (map_det) {\color{white}{\footnotesize SWH Det.}};
	\node [above of=map_det,node distance=0cm,yshift=0.2cm] () {\footnotesize SISO};
	\node [above of=map_det,node distance=0cm,yshift=-0.2cm] () {\footnotesize Detector};
	
	\node [right of=map_det,node distance=1.8cm,yshift=-0cm] (mapper_dummy){};
	\node [right of=map_det, node distance=0.6cm,yshift=0.27cm] (demapper) {};
	\node [right of=map_det, node distance=0.6cm,yshift=-0.27cm] (mapper) {};
	\node [above of=mapper,node distance=0cm,yshift=-0	cm] () {};

	\node [block, right of=map_det, node distance=3.25cm] (SISO_dec2) {\color{white}{\footnotesize SWH Det}};
	\node [above of=SISO_dec2,node distance=0cm,yshift=0.2cm] () {\footnotesize SISO};
	\node [above of=SISO_dec2,node distance=0cm,yshift=-0.2cm] () {\footnotesize Decoder};
	
	\node [output, name=output2, right of=SISO_dec2,node distance=1.2cm] {};

	\node [right of=demapper,node distance=0.5cm,yshift=0.21cm] () {\footnotesize $\mathbf{L}^{\rm e}$};
	\node [right of=mapper,node distance=0.5cm,yshift=-0.21cm] () {\footnotesize $\mathbf{L}^{\rm a}$};

	\draw [->] (fft) -- node [xshift=-0.1cm]{\footnotesize $\mathbf{y}$} (map_det);	
	
	\draw [->] (SISO_dec2) -- node [xshift=0cm]{\footnotesize $\hat{\mathbf{b}}$} (output2);
	
	\node [block2, right of=demapper, node distance=1.2cm,yshift=0cm] (deinterleaver) {\color{black}{\footnotesize $\Pi^{-1}$}};
	\node [block2, right of=mapper, node distance=1.2cm,yshift=-0cm] (interleaver) {\color{black}{\footnotesize $\Pi$}};
	\node [right of=deinterleaver, node distance=0.9cm,yshift=0cm] (deinterleaver_dummy) {};
	\node [right of=interleaver, node distance=0.9cm,yshift=0cm] (interleaver_dummy) {};
	
	\draw [->] [](demapper) -- node []{}(deinterleaver);
	\draw [->] [](interleaver) -- node []{}(mapper);
	
	\draw [->] [](deinterleaver) -- node []{}(deinterleaver_dummy);
	\draw [->] [](interleaver_dummy) -- node []{}(interleaver);
	
	\node [left of=channel, node distance=0.45cm,yshift=1cm] (tx_top) {};
	\node [left of=channel, node distance=0.45cm,yshift=-0.8cm] (tx_bottom) {};
	\draw [-,dashed] (tx_top) -- node []{} (tx_bottom);	
	
	\node [left of=fft, node distance=0.45cm,yshift=1cm] (tx_top) {};
	\node [left of=fft, node distance=0.45cm,yshift=-0.8cm] (tx_bottom) {};
	\draw [-,dashed] (tx_top) -- node []{} (tx_bottom);	
	
	\node [] at (5,1) {\footnotesize transmitter};
	\node [] at (10.5,1) {\footnotesize channel};
	\node [] at (15,1) {\footnotesize receiver};
	
\end{tikzpicture} 
	\vspace{-1.2cm}
	\caption{\textcolor{black}{Block diagram of the system model including transmitter, channel and receiver.}}
	\label{fig:system_model}
\end{figure*}

The transmitter employs a bit-interleaved coded modulation (BICM) transmission \cite{BICM}. As depicted in \textcolor{black}{the transmitter part} of Fig.~\ref{fig:system_model}, a vector of information bits $\ma{b} \in \left\{0,1\right\}^{N_{\rm b}}$ is encoded creating the coded bit stream $\ma{c}' = {\rm enc}\pa{\ma{b}} \in \left\{0,1\right\}^{N_{\rm c}}$, where ${\rm enc}(\cdot)$ represents the encoder.
The quantities $R = {N_{\rm b}}/{N_{\rm c}}$, $N_{\rm b}$ and $N_{\rm c}$ represent the coding rate, number of uncoded and coded bits, respectively. 
Subsequently, the coded bits $\ma{c}$ are interleaved as $\ma{c} = \Pi\pa{\ma{c}'}$, where $\Pi (\cdot)$ represents the interleaver.
As shown in Fig.~\ref{fig:system_model}, the vector of coded bits is split into two sub-vectors ${\ma{c}}^{\rm I} = [ {c_0 \,\, c_1 \,\, \cdots \,\, c_{\frac{N_{\rm c}}{2}-1}} ]^{\rm T},{\ma{c}}^{\rm Q} = [ {c_{\frac{N_{\rm c}}{2}} \,\, c_{\frac{N_{\rm c}}{2}+1} \,\, \cdots \,\, c_{N_{\rm c}-1}} ]^{\rm T} \in \left\{0,1\right\}^{\frac{N_{\rm c}}{2}}$, which are mapped respectively onto the in-phase and quadrature \ac{PAM} symbols $\mathbf{r}^{\rm I},\mathbf{r}^{\rm Q} \in \mathcal{R}^N$, where $\mathcal{R}$ denotes the set of a \ac{PAM} constellation with cardinality $|\mathcal{R}|=\sqrt{J}$.  
The \ac{QAM} symbols can be defined as ${\mathbf{d}} = {\mathbf{r}}^{\rm I} + j{\mathbf{r}}^{\rm Q} \in \mathcal{D}^{N}$, where $\mathcal{D}$ is the QAM set with cardinality $|\mathcal{D}| = J$.

In general, the QAM symbols are modulated with the FD precoding matrix $\mathbf{A} \in \mathbb{C}^{N \times N}$ as
\begin{equation}\label{eq:x}
	\ma{x} = \ma{A}\ma{d},
\end{equation}
\textcolor{black}{where the modulated signal in the time domain is obtained by performing the inverse \ac{FFT} as $\tilde{\mathbf{x}} = \ma{F}_N^{\rm H}\ma{x}$.}
In this work, we consider three options for the matrix $\mathbf{A}$.
\subsubsection{\ac{SWH}} $\ma{A}_{\rm SW} = \ma{W}_Q \otimes \ma{I}_P$\footnote{In this paper we neglect the parameter $Q'$ of \cite{BomfinSWH} because we consider a time invariant channel model.}, $QP = N$, $\mathbf{W}_Q$ is the normalized Walsh-Hadamard matrix of size $Q$, 
\begin{equation}
    \ma{W}_Q = \ma{W}_{2} \otimes \ma{W}_{{Q}/{2}}, \,\,\,\, \ma{W}_{2} = \frac{1}{\sqrt{2}}\begin{bmatrix}
1 & \phantom{-}1\\ 
1 & -1
\end{bmatrix} 
\end{equation}
and $\otimes$ is the Kronecker product. 
As it has been shown in \cite{BomfinSWH}, the precoder $\ma{A}$ ensures equal channel gain for the elements of $\ma{d}$ for $Q \geq L$, where $L$ is the channel length.
A detailed description of the sparse spreading precoder is given in Section \ref{sec:sparse_precoders}.

\subsubsection{\ac{DFT}} $\ma{A}_{\rm F} = \mathbf{F}_N$ is defined as the DFT matrix of size $N$. Notice that this is the same as the DFT-spread modulation, sometimes also referred to as single-carrier.

\subsubsection{\ac{SDFT}} $\ma{A}_{\rm SF} = \ma{F}_Q \otimes \ma{I}_P$. This modulation follows the same principle as the SWH of \cite{BomfinSWH}, but with the WHT being replaced by the DFT.

\subsection{Received Signal}
\textcolor{black}{Assuming perfect time and frequency synchronization, and proper \ac{CP} insertion/removal, the discrete-time received signal in the time domain $\tilde{\ma{y}} \in \mathbb{C}^N$ is given by
\begin{equation}
	\tilde{\mathbf{y}} = \mathbf{H}\tilde{\mathbf{x}} + \tilde{\mathbf{w}},
\end{equation}	
where $\mathbf{H}\in \mathbb{C}^{N\times N}$ is a circulant matrix whose first column is defined by the channel impulse response $\ma{h}$ and $\tilde{\ma{w}} \sim \mathcal{CN}\left( 0,\ma{I}_N \sigma^2 \right)$ is the \ac{AWGN} with power $\sigma^2$.
\textcolor{black}{The above signal is depicted in the channel part in Fig.~\ref{fig:system_model}}.
In the frequency domain, the received signal is given in double column version
%
%
\begin{equation}\label{eq:y}
	\begin{split}
	\ma{y} & = \ma{F}_N	\tilde{\mathbf{y}}\\  
			& = \ma{\Lambda}\ma{A}\ma{d} + \ma{w},
	\end{split}
\end{equation}
where $\ma{\Lambda} = \ma{F}_N	\ma{H} \ma{F}_N^{\rm H}\in \mathbb{C}^{N \times N}$ is the diagonal channel matrix whose elements correspond to the channel response in the frequency domain.
The component $\ma{w} \sim \mathcal{CN}\left( 0,\ma{I}_N \sigma^2 \right)$ is the \ac{AWGN} with power $\sigma^2$ in frequency domain.
The second line of \eqref{eq:y} is obtained by substituting $\tilde{\mathbf{x}} = \ma{F}_N^{\rm H}\ma{x} =  \ma{F}_N^{\rm H}\ma{Ad}$ from~\eqref{eq:x}.}
\subsection{Iterative Receiver}\label{subsec:iterative_receiver}
%
%
A generic structure of the iterative receiver is depicted in \textcolor{black}{the right most part of} Fig.~\ref{fig:system_model}, where perfect channel knowledge and noise variance are known\footnote{\textcolor{black}{We note that in case of imperfect channel knowledge, the system in \eqref{eq:y} could be modeled with an additional noise term w.r.t. to the channel estimation error as in \cite{BomfinTWC}, which in turn changes the SNR operating point of the system, and does not fundamentally change the design of the equalizer.}}.
There are basically two main structures that compose this receiver, namely, the \ac{SISO} detector and the SISO decoder.

In this work, we focus on the SISO detector which has the purpose of estimating extrinsic bit \acp{LLR}, $\mathbf{L}^{\rm e}$, given the received signal, channel response, noise power and a-priori bit LLRs, $\mathbf{L}^{\rm a}$. 
The terms extrinsic and a-apriori LLRs are defined from the detector's perspective.
In this work, we consider iterative receivers with two classes of detectors. 
The first class is based on the MAP detector for the SWH which is described in Section~\ref{sec:map_detectors_SWH}.
The second class of detectors is the \ac{SILE-EPIC} of \cite{Sahin}, which is an EP-based receiver with one-tap FD equalization.
This receiver performs self-iterations between the LMMSE equalizer and the mapper/demapper nodes based on the \ac{EP} method, which drastically improves the performance under highly selective channels.

%
The other component is the SISO decoder, which updates the coded bit LLRs to the detector node based on its input bit LLRs. 
In order to respect the turbo principle of not using the same information twice, $\mathbf{L}^{\rm a}$ is computed by subtracting  $\mathbf{L}^{\rm e}$ from the decoder's \ac{APP}  estimation.
In this work, we employ the \ac{RSC} encoder with \ac{BCJR} SISO decoding.

\section{Sparse Precoders System Model}\label{sec:sparse_precoders}
In this section, we present the system model for SWH and SDFT precoders, which is necessary to describe their respective receivers.
In the following, we consider the SWH precoder $\ma{A}_{\rm SW} = \ma{W}_Q \otimes \ma{I}_P$, but the model is directly applicable to the DFT precoder by replacing $\ma{W}_Q$ by $\ma{F}_Q$ as mentioned in Section~\ref{subsec:transmitter}.

\subsection{Modulation}
One way to describe the sparsity of $\ma{A}_{\rm SW}$ is to group $P$ different sub-vectors with $Q$ data symbols each. 
The data symbols of each sub-vector are spread over the same set of sub-carriers.
These sub-vectors are defined as ${\mathbf{d}_p} = [ {{d}}_{p} \,\, {{d}}_{p+P} \,\, {{d}}_{p+2P} \,\, \cdots \,\, {{d}}_{p+P(Q-1)}]^{\rm T} \in \mathcal{D}^{Q}$ for $p = 0,1,\cdots,P-1$.
It is possible to show that the symbols of ${\mathbf{d}}_p$ are spread among the carriers with the same index as the symbols as
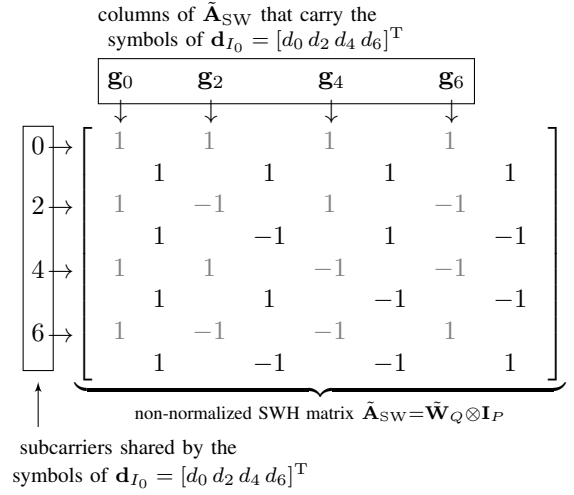
\begin{figure}[t!]
	\tikzstyle{block} = [draw, fill=white, rectangle, 
    minimum height=3em, minimum width=3em]
\tikzstyle{block2} = [draw, fill=white, rectangle, 
minimum height=1em, minimum width=2.4em]
    
\tikzstyle{multiplier} = [draw,circle,fill=blue!20,add={}{}{}{}] {} 
\tikzstyle{sum} = [draw,circle,fill=blue!20,add2={}{}{}{}] {} 
\tikzstyle{input} = [coordinate]
\tikzstyle{output} = [coordinate]
\tikzstyle{pinstyle} = [pin edge={to-,thin,black}]

\def\windup{
	\tikz[remember picture,overlay]{
		\draw (-0.8,0) -- (0.8,0);
		\draw (-0.8,-0.4)--(-0.4,-0.4) -- (0.4,0.4) --(0.8,0.4);
}}

\centering
\begin{tikzpicture}[auto, node distance=2cm,>=latex']

\draw[draw=black] (-2.5,2.3) rectangle ++(5,0.6);

\draw[draw=black] (-3.5,2) rectangle ++(0.4,-3.5+0.25);

\node [input, name=input2] at (0,0) {};

\node [] at (0.45,0) {$\underbrace{\begin{bmatrix}
{\color{white}-} {\color{gray}1} &  &  {\color{gray}1} &  &  {\color{gray}1} &  &  {\color{gray}1} &  \\ 
&  1&  &  1 &  &  1 &  & 1{\color{white}-} \\ 
{\color{white}-}{\color{gray}1} &  &  {\color{gray}-1}&  &  {\color{gray}1} &  &  {\color{gray}-1}&  \\ 
&  1&  &  -1&   &  1 &  & -1{\color{white}-}\\ 
{\color{white}-}{\color{gray}1} &  &  {\color{gray}1} &  &  {\color{gray}-1}&  &  {\color{gray}-1}&  \\ 
&  1&  &  1 &  &  -1&  & -1{\color{white}-}\\ 
{\color{white}-}{\color{gray}1} &  &  {\color{gray}-1}&  &  {\color{gray}-1}&  &  {\color{gray}1} &  \\ 
&  1&  &  -1&  &  -1&  & 1 {\color{white}-}
\end{bmatrix}}_{{\text{non-normalized SWH matrix}} \,\, \tilde{\mathbf{A}}_{\rm SW}=\tilde{\ma{W}}_Q \otimes \ma{I}_P}$};

\node [] at (-2.2,2.4+0.2) {$\mathbf{g}_0$};
\node [] at (-2.2,2+0.2) {$\downarrow$};

\node [] at (-1,2.4+0.2) {$\mathbf{g}_2$};
\node [] at (-1,2+0.2) {$\downarrow$};

\node [] at (0.6,2.4+0.2) {$\mathbf{g}_4$};
\node [] at (0.6,2+0.2) {$\downarrow$};

\node [] at (2.2,2.4+0.2) {$\mathbf{g}_6$};
\node [] at (2.2,2+0.2) {$\downarrow$};

\node [] at (-3.3,1.5+0.25) {$0$};
\node [] at (-3,1.45+0.25) {$\rightarrow$};

\node [] at (-3.3,1.5-0.8+0.25) {$2$};
\node [] at (-3,1.45-0.8+0.25) {$\rightarrow$};

\node [] at (-3.3,1.5-1.65+0.25) {$4$};
\node [] at (-3,1.45-1.65+0.25) {$\rightarrow$};

\node [] at (-3.3,1.5-2.5+0.25) {$6$};
\node [] at (-3,1.45-2.5+0.25) {$\rightarrow$};

\node [] at (-0.65,3.5) {\footnotesize columns of $\tilde{\ma{A}}_{\rm SW}$ that carry the};
\node [] at (-0.4,3.15) {\footnotesize symbols of ${\mathbf{d}}_{I_0} = [d_0 \, d_2\,d_4\,d_6]^{\rm T}$};

\node [] at (-1.7-0.4,-2.3) {\footnotesize subcarriers shared by the };
\node [] at (-1.27-0.4,-2.65) {\footnotesize symbols of  ${\mathbf{d}}_{I_0}=[d_0 \, d_2\,d_4\,d_6]^{\rm T}$};

\draw [->] (-3.3,-2) -- node []{} (-3.3,-1.4);

\end{tikzpicture}
	\vspace{-0.8cm}
	\caption{Example of $\ma{A}_{\rm SW}$: $N=8$, $Q=4$ and $P=2$, where the elements of ${\mathbf{d}}_0 = [d_0 \, d_2\,d_4\,d_6]^{\rm T}$ share the same sub-carriers with indexes $I_0 \in \left\{ 0,2,4,6\right\}$. The zeros of the SWH matrix $\tilde{\ma{A}}=\tilde{\ma{W}}_Q \otimes \ma{I}_P$ are replaced with spaces for clearness and $\ma{g}_n$ is the $n$th columns of $\tilde{\ma{A}}_{\rm SW}$.}
	\label{fig:SWH_proof}
\end{figure}
\begin{equation}\label{eq:Ip}
	I_p = \left\{p+qP| q\in\left\{0,1, \cdots, Q-1 \right\}\right\}.
\end{equation}
This is illustrated in Fig.~\ref{fig:SWH_proof} by depicting the non-normalized matrix $\tilde{\ma{A}}_{\rm SW}$ with $N=8$ sub-carriers, $Q=4$ symbols per sub-group and $P = 2$ groups.
Notice that the columns and rows of $\ma{A}_{\rm SW}$ are associated with the symbols and sub-carriers, respectively.
The symbol sub-vectors are defined as ${\mathbf{d}}_0 = [d_0 \, d_2 \, d_4 \, d_6]^{\rm T}$ and ${\mathbf{d}}_1 = [d_1 \, d_3\,d_5\,d_7]^{\rm T}$ for $p=0$ and $p=1$, respectively. 
Then, one can verify that the modulation $\ma{A}_{\rm SW}\ma{d}$ in \eqref{eq:x} spreads the symbols with indexes $I_0 \in \left\{ 0,2,4,6\right\}$ among the carriers with the same indexes $I_0$. 

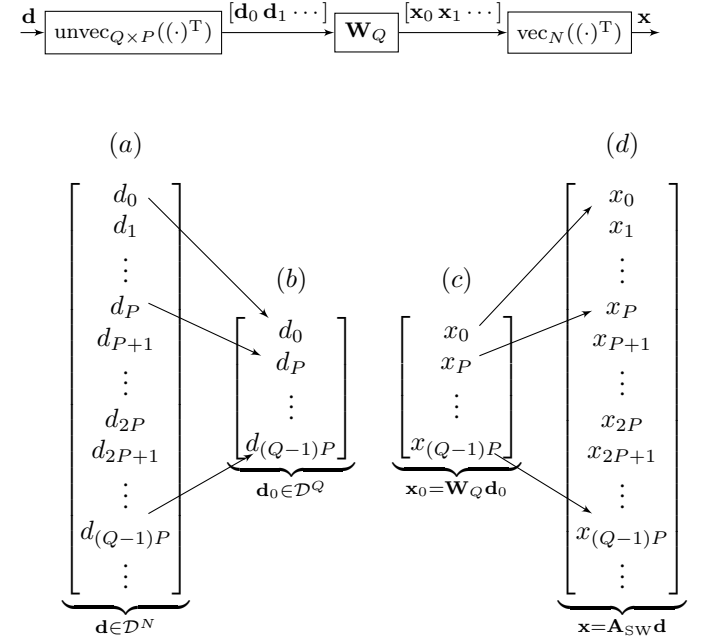
\begin{figure}[t!]
	\tikzstyle{block} = [draw, fill=white, rectangle, 
minimum height=3em, minimum width=3em]
\tikzstyle{block2} = [draw, fill=white, rectangle, 
minimum height=1em, minimum width=2.4em]

\tikzstyle{multiplier} = [draw,circle,fill=blue!20,add={}{}{}{}] {} 
\tikzstyle{sum} = [draw,circle,fill=blue!20,add2={}{}{}{}] {} 
\tikzstyle{input} = [coordinate]
\tikzstyle{output} = [coordinate]
\tikzstyle{pinstyle} = [pin edge={to-,thin,black}]

\def\windup{
	\tikz[remember picture,overlay]{
		\draw (-0.8,0) -- (0.8,0);
		\draw (-0.8,-0.4)--(-0.4,-0.4) -- (0.4,0.4) --(0.8,0.4);
}}

\centering
\begin{tikzpicture}[auto, node distance=2cm,>=latex']

		\node [input, name=input2] at (-1.4,5) {};
	
	\node [block2, right of=input2,node distance=1.5cm] (unvec) {\color{black}{\footnotesize ${\rm unvec}_{Q \times P}((\cdot)^{\rm T})$}};
	\node [block2, right of=unvec,node distance=3.1cm] (modulator) {\color{black}{\footnotesize $\mathbf{W}_Q$}};
	\node [block2, right of=modulator,node distance=2.7cm] (vec) {\color{black}{\footnotesize ${\rm vec}_{N}((\cdot)^{\rm T})$}};
	
	\node [input, right of=vec,node distance=1.2cm] (output) {};
	
	\draw [->] (input2) -- node [xshift=-0.05cm]{\footnotesize $\mathbf{d}$} (unvec);
	\draw [->] (unvec) -- node [xshift=0cm]{\footnotesize $[\ma{d}_0 \, \ma{d}_1 \cdots]$} (modulator);
	\draw [->] (modulator) -- node [xshift=-0cm]{\footnotesize $[\ma{x}_0 \, \ma{x}_1 \cdots]$} (vec);
	\draw [->] (vec) -- node [xshift=-0.02cm]{\footnotesize $\ma{x}$} (output);

\node [] at (0,0) {$\underbrace{\begin{bmatrix}
		{d_0} \\
		d_1 \\
		\vdots \\
		d_{P} \\
		d_{P+1} \\
		\vdots \\
		d_{2P} \\
		d_{2P+1} \\
		\vdots \\
		d_{(Q-1)P} \\
		\vdots
\end{bmatrix}}_{\mathbf{d}\in \mathcal{D}^N}$};

\node [] at (2.2,0) {$\underbrace{\begin{bmatrix}
	d_0 \\
	d_P \\
	\vdots \\
	d_{(Q-1)P}
\end{bmatrix}}_{\mathbf{d}_0 \in \mathcal{D}^Q}$};

\node [] at (4.4,0) {$\underbrace{\begin{bmatrix}
			x_0 \\
			x_P \\
			\vdots \\
			x_{(Q-1)P}
\end{bmatrix}}_{\mathbf{x}_0 = \mathbf{W}_Q\mathbf{d}_0}$};

\node [] at (6.6,0) {$\underbrace{\begin{bmatrix}
			x_0 \\
			x_1 \\
			\vdots \\
			x_{P} \\
			x_{P+1} \\
			\vdots \\
			x_{2P} \\
			x_{2P+1} \\
			\vdots \\
			x_{(Q-1)P} \\
			\vdots
\end{bmatrix}}_{\mathbf{x} = \ma{A}_{\rm SW}\ma{d}}$};

	\draw [->] (0.3,2.8) -- node []{} (1.9,1.2);
	\draw [->] (0.3,1.4) -- node []{} (1.8,0.7);
	\draw [->] (0.3,-1.4) -- node []{} (1.7,-0.6);
	
	\draw [->] (4.7,1.1) -- node []{} (6.2,2.7);
	\draw [->] (4.7,0.7) -- node []{} (6.2,1.3);
	\draw [->] (4.9,-0.6) -- node []{} (6.2,-1.4);
	
	\node [] at (0,3.5) {$(a)$};
	\node [] at (2.2,1.7) {$(b)$};
	\node [] at (4.4,1.7) {$(c)$};
	\node [] at (6.6,3.5) {$(d)$};

\end{tikzpicture}
	\vspace{-0.8cm}
	\caption{(Top) Sparse modulation diagram $\ma{x} = \ma{A}_{\rm SW}\ma{d}$. (Bottom) Representation of $\rm unvec$ and $\rm vec$ operations for $p=0$, $(a)$ data vector, $(b)$ data sub-vector after $\rm unvec$ operation, $(c)$ modulation per sub-vector, $(d)$ sparse modulated data after $\rm vec$ operation.}
	\label{fig:sparse_spreading}
\end{figure}

Additionally, we can define the sparse modulation as in Fig.~\ref{fig:sparse_spreading} which is compactly written as
\begin{equation}
	\ma{x} = {\rm vec}_N\left(\left(\ma{W}_Q \left({\rm unvec}_{Q \times P} (\ma{d^{\rm T}} )\right)\right)^{\rm T}\right).
\end{equation}
The modulated data is computed in three steps, namely, i) stack the sub-vectors $[\ma{d}_0 \, \ma{d}_1 \cdots]$ with the operation ${\rm unvec}_{Q \times P}(\ma{d}^{\rm T})$, ii) modulate the sub-vectors as $[\ma{x}_0 \, \ma{x}_1 \cdots] = \ma{W}_Q[\ma{d}_0 \, \ma{d}_1 \cdots]$, and iii) create an $N \times 1$ vector representing the modulated signal as  $\ma{x} = {\rm vec}_N([\ma{x}_0 \, \ma{x}_1 \cdots]^{\rm T})$.

\subsection{Received Signal}
Since the received signal in the equation \eqref{eq:y} is already in FD, we can build its sub-vectors in analogy to ${\mathbf{d}}_p$ as $\mathbf{y}_p = [ {{y}}_{p} \,\, {{y}}_{p+P} \cdots \,\, {{y}}_{p+P(Q-1)}]^{\rm T} \in \mathbb{C}^{Q \times 1}$ for $p = 0,1,\cdots,P-1$, which is modeled as
\begin{equation}\label{eq:yp}
		\mathbf{y}_p  = {\mathbf{\Lambda}}_p\mathbf{W}_Q\mathbf{d}_p+ {\mathbf{w}}_p.
\end{equation}
The channel matrix and noise can be split accordingly as ${\mathbf{\Lambda}}_{p} = {\rm diag}([ {{\Lambda}}_{p,p} \,\, {{\Lambda}}_{p+P,p+P} \,\, \cdots {{\Lambda}}_{p+P(Q-1),p+P(Q-1)}])$ $\in \mathbb{C}^{Q \times Q}$ has the channel components in its diagonal whose indexes belong to $I_p$, and ${\mathbf{w}}_p =  [ {{w}}_{p} \,\, {{w}}_{p+P} \cdots {{w}}_{p+P(Q-1)}]^{\rm T}\sim \mathcal{N}\left( 0,\ma{I}_Q\sigma^2 \right)$, respectively.

\section{MAP Detectors for the SWH}\label{sec:map_detectors_SWH}
%
%
\subsection{Channel Phase Correction}
As shown in \cite{BomfinMAP}, in order to build an equalizer with independent in-phase and quadrature processing, the phase of the received signal $\ma{y}_p$ in the equation \eqref{eq:yp} is corrected such that its real and imaginary components are exclusively associated with the in-phase and quadrature dimensions of the QAM constellation, respectively.
This can be done by correcting the channel phase rotation as
\begin{equation}\label{eq:Y_pp}
	\begin{split}
		{\mathbf{y}'_p} = \frac{{\mathbf{\Lambda}_p}^{\rm H}}{|{\mathbf{\Lambda}_p}|}{\mathbf{y}_p} = |{\mathbf{\Lambda}_p}|\mathbf{W}_Q({\mathbf{r}^{\rm I}_p + j\mathbf{r}^{\rm Q}_p})+ {\mathbf{v}'_p},
	\end{split}
\end{equation}
where ${\mathbf{v}}'_p\sim \mathcal{CN}\left( 0,\ma{I}_Q \sigma^2 \right)$ is the noise with phase rotation and has its statistics unchanged in relation to ${\mathbf{v}_p}$ in \eqref{eq:yp}.
Also, $|{\mathbf{\Lambda}_p}|$ returns the element-wise absolute values of ${\mathbf{\Lambda}_p}$.
Since the modulation matrix $\mathbf{W}_Q$ does not introduce any phase rotation in the data symbols, it is straightforward to observe that $	\mathbf{y}_p^{{\rm I}}={\rm Re}({\mathbf{y}}_p')$ and $	\mathbf{y}_p^{{\rm Q}}={\rm Im}({\mathbf{y}_p'})$  exclusively carry information of the in-phase and quadrature components associated to the PAM constellation, respectively, as shown below
\begin{equation}\label{eq:yI_p}
	\begin{split}	
		\mathbf{y}^{{\rm I}}_p  = |{\mathbf{\Lambda}}_p|\mathbf{W}_Q\mathbf{r}^{\rm I}_p+ {\mathbf{w}}^{{\rm I}}_p, \,\,\,\,\,\,\,\,\,\,
		\mathbf{y}^{{\rm Q}}_p =  |{\mathbf{\Lambda}}_p|\mathbf{W}_Q\mathbf{r}^{\rm Q}_p+ {\mathbf{w}}^{{\rm Q}}_p
	\end{split}.
\end{equation}
Thus, the MAP-based receivers for the SWH precoder operate directly on the received signal defined in \eqref{eq:yI_p}.

\subsection{Exact MAP Detector \cite{BomfinMAP}}\label{subsec:exat_app_detector}
In the following, the detector estimation is described for the symbols and bits associated with the in-phase PAM symbols without loss of generality, since the same analysis applies to the quadrature component.
As shown in Appendix~\ref{appendix:MAP}, we can write the conditional probability below
\begin{equation}\label{eq:p_np}
	\begin{split}
		p({\mathbf{y}}^{\rm I}_{p}|{\rm\mathbf{z}}) & \propto \exp\left(-\frac{1}{\sigma^2}\left \| {\mathbf{y}}^{\rm I}_{p}-|{\mathbf{\Lambda}}_p|\mathbf{W}_{Q}{\rm\mathbf{z}} \right \|^2 \right)
		\\ & = \exp\left(\sum_{\substack{q'=0, \\ i = \mathcal{S}(\ma{z},q)}}^{Q-1} C^{\rm I}_{p}(q',i)\right),
	\end{split}
\end{equation}
\textcolor{black}{``for a given PAM vector realization $\mathbf{z} \in \mathcal{R}^Q$.}
The quantity $C^{\rm I}_{p}(q,i)$ is given by
\textcolor{black}{\begin{equation}\label{eq:C}
	C^{\rm I}_{p}(q,i) = -(\ma{y}_p^{\rm I}[q] - |\ma{\Lambda}_{p}[q,q]|s_i)^2/\sigma^2,
\end{equation}}
for $q \in\left\{ 0,1,\cdots,Q-1\right\}$ and $i \in\left\{ 0,1,\cdots,Q(\sqrt{J}\!-\!1)\right\}$, where
\begin{equation}
	s_i = (-Q(M\!-\!1) \!+ \!2i)/\sqrt{Q}
\end{equation}
is based on the set $\mathcal{U}$ defined in Appendix~\ref{appendix:MAP}.
The advantage defining $C^{\rm I}_{p}(q,i)$ in \eqref{eq:C} is that it allows the computation of $p({\mathbf{y}}^{\rm I}_{p}|{\rm\mathbf{z}})$ in a simplified manner as the second line of \eqref{eq:p_np}.
\textcolor{black}{We note that $C^{\rm I}_{p}(q,i)$ should be computed for each channel realization, however, since $i \in\left\{ 0,1,\cdots,Q(\sqrt{J}\!-\!1)\right\}$, there are $Q(\sqrt{J}\!-\!1)+1$ values to be computed per PAM symbol, which does not represent a bottleneck in terms of complexity. 
More details are given in Section \ref{sec:complexity_analysis}.}
%
%
%
%
%
In \eqref{eq:p_np}, {$\mathcal{S}(\ma{z},q) = \{i | \textcolor{black}{u_q = s_i}\}$} returns the index $i$ which attains the condition \textcolor{black}{$u_q = s_i$} and can be interpreted as a database which stores the respective values for all $\ma{z} \in \mathcal{R}^Q$ and $q \in\left\{ 0,1,\cdots,Q-1\right\}$, \textcolor{black}{where $\ma{u} = \ma{W}_Q\ma{z}$.}

In summary, the values $C^{\rm I}_{p}(q,i)$ are computed beforehand and then reused for each new $\ma{z}$, which allows the receiver to avoid most of the complex multiplications in relation to the general MAP formulation. 
On the contrary, if the quantity $\left \| {\mathbf{y}}^{\rm I}_{p}-|{\mathbf{\Lambda}}_p|\mathbf{W}_{Q}{\rm\mathbf{z}} \right \|^2$ is computed for each $\ma{z}$ in its general form in \eqref{eq:p_np}, the calculations of \eqref{eq:C} are unnecessarily computed multiple times because the possible values of the elements of the modulated symbols $\mathbf{W}_{Q}{\rm\mathbf{z}}$ are taken from a small set.

In general, the MAP estimation of the in-phase or quadrature PAM data symbols is computed as
\begin{equation}\label{eq:p_np2}
	\begin{split}
		p_{p,q}^{\rm I}({\rm r}) & = p(r^{\rm I}_{p,q}= {\rm r} |{\mathbf{y}}^{\rm I}_{p})
		\\& \propto \sum_{\substack{{\rm \mathbf{z}} \in \mathcal{R}^{Q}, \\{\rm z}_q={\rm r}}} p({\mathbf{y}}^{\rm I}_{p}|\mathbf{z})p^{\rm a,I}_{p}({\mathbf{r}}^{\rm I}_p={\rm \mathbf{z}}),
	\end{split}
\end{equation}
for $q\in \left\{0,1,\cdots, Q-1\right\}$, $p \in \left\{0,1,\cdots, P-1 \right\}$ where $	p({\mathbf{y}}^{\rm I}_{p}|{\rm\mathbf{z}})$ is given in \eqref{eq:p_np}
and the a-priori symbol probability under the assumption of independent bits is given by
\begin{equation}\label{eq:pa}
	p^{\rm a,I}_{p}(\mathbf{r}^{\rm I}_p={\rm \mathbf{z}}) = \prod_{\substack{q = 0}}^{Q-1}\prod_{b=0}^{\frac{1}{2}\log_2 \! J-1}  \exp\left(\phi_b^{-1}({\rm z}_{q}) L^{{\rm a,I}}_{p,q,b}\right),
\end{equation}
where $\phi_b^{-1}({\rm z})$ is a demapper which returns $1$ or $0$ if the $b$th bit of the PAM symbol ${\rm z} \in \mathcal{R}$ is 0 or 1, respectively.
For completeness, the indexes $p$, $q$ and $b$ of $L^{{\rm a,I}}_{p,q,b}$ correspond to the $p$th sub-vector, $q$th symbol and $b$th bit.
The posterior symbol probability in \eqref{eq:p_np2} can be written as \eqref{eq:p_r} by combining \eqref{eq:p_np} with \eqref{eq:pa}.
Lastly, the extrinsic LLRs are computed by combining \eqref{eq:p_r} with
\begin{figure*}[!t]
	\normalsize
	\setcounter{MYtempeqncnt}{\value{equation}}
	\begin{equation}
		\label{eq:p_r}
		p_{p,q}^{\rm I}({\rm r}) \propto \sum_{\substack{{\rm \mathbf{z}} \in \mathcal{R}^{Q}, \\{\rm z}_q={\rm r}}} \exp\left(\sum_{\substack{q' = 0\\ i = \mathcal{S}(\ma{z},q')}}^{Q-1} \left(C^{\rm I}_{p}(q',i) - \sum_{b=0}^{\frac{1}{2}\log_2 (J)-1}  \phi_b^{-1}({\rm z}_{q'}) L^{{\rm a,I}}_{p,q',b}  \right) \right)
	\end{equation}
	\setcounter{equation}{\value{MYtempeqncnt}+1}
	\hrulefill
	\vspace*{4pt}
\end{figure*}
\begin{equation}\label{eq:Le}
	L^{{\rm e,I}}_{p,q,b} = \ln{\frac{\sum_{{\rm r} \in \mathcal{R}_{b}^{(0)}} p^{\rm I}_{p,q}({\rm r}) }{\sum_{{\rm r} \in \mathcal{R}_{b}^{(1)}} p_{p,q}^{\rm I}({\rm r})}}-L^{{\rm a,I}}_{p,q,b},
\end{equation}
where $\mathcal{R}_{b}^{(0)} = \{{\rm z} \in \mathcal{R} | \phi_b({\rm z})^{-1} = 0 \}$ is the set containing all PAM symbols whose $b$th is 0, and $\mathcal{R}_{b}^{(1)}$ is the equivalent set for the bit 1.

The issue of computing the LogSumExp in \eqref{eq:Le} in combination with \eqref{eq:p_r} is not discussed in \cite{BomfinMAP}.
Thus, in the following we provide two \textcolor{black}{feasible} approaches to tackle this problem, namely, the Log-MAP and Max-Log-MAP \cite{Robertson} methods.

\subsection{Log-MAP Detector}\label{subsec:logMAP_det}
For the MAP-Log detector, we are not interested in the intermediate symbol marginalization in \eqref{eq:p_r}, but in computing the extrinsic LLRs directly. 
In order to allow this computation, the following variable is defined
\begin{equation}\label{eq:t_pq}
	\begin{split}
		t^{\rm I}_{p}(\ma{z}) =  \sum_{\substack{q' = 0\\ i = \mathcal{S}(\ma{z},q')}}^{Q-1} \left(C_{p}(q',i) - \sum_{b=0}^{\frac{1}{2}\log_2 \textcolor{black}{(J)}-1}  \phi_b^{-1}({\rm z}_{q'}) L^{{\rm a,I}}_{p,q',b}  \right),
	\end{split}
\end{equation}
which is essentially the exponential argument in the equation \eqref{eq:p_r} for a given $\ma{z} \in \mathcal{R}^Q$.
Then, the extrinsic LLRs of \eqref{eq:Le} are written as a function of $t^{\rm I}_{p}(\ma{z})$ as follows
\begin{equation}\label{eq:Le2}
		\begin{split}
		L^{{\rm e,I}}_{p,q,b} = & \ln\left(\sum_{\ma{z} \in \mathcal{R}^Q_{q,b,0}}\exp \left(t^{\rm I}_{p}(\ma{z})\right)\right)
	\\ &	-  \ln\left(\sum_{\ma{z} \in \mathcal{R}^Q_{q,b,1}}\exp \left(t^{\rm I}_{p}(\ma{z})\right)\right)-L^{{\rm a,I}}_{p,q,b},
	\end{split}
\end{equation}
where the set $\mathcal{R}^Q_{q,b,0} = \{ \ma{z} \in \mathcal{R}^Q | \phi_b({\rm z}_q)^{-1} = 0\}$ contains all realizations of $\ma{z}$ in which the $b$th bit of ${\rm z}_q$ equals 0, and $\mathcal{R}^Q_{q,b,1}$ has the same definition for the bit 1. 

When computed in closed form, the Log-MAP method provides an exact computation of the MAP detector based on the LogSumExp relation
\begin{equation}\label{eq:log_sum}
	\ln(e^{\delta_0}+e^{\delta_1}) = \max(\delta_0,\delta_1) + \ln(1+e^{-|\delta_0 - \delta_1|^2}).
\end{equation}
If we write $e^\delta = e^{\delta_0}+e^{\delta_1}$ , we can compute $\ln(e^{\delta_0} + e^{\delta_1} + e^{\delta_2}) = \ln(e^\delta + e^{\delta_2}) = \max(\delta,\delta_2) + \ln(1+e^{-|\delta - \delta_2|^2})$ using \eqref{eq:log_sum}.
Since this process always accepts a new term to the sum of exponential \cite{Robertson}, the desired expression $\ln(e^{\delta_0}+e^{\delta_1}+ \cdots e^{\delta_{N-1}})$ can be computed exactly in \eqref{eq:Le2}.

Notice that the term $f_c(|\delta_0 - \delta_1|) = \ln(1+e^{-|\delta_0 - \delta_1|^2})$ in \eqref{eq:log_sum} is still not friendly for practical implementation.
Thus, one solution is to store the values of $f_c(\cdot)$ in a table, such that $\ln(\cdot)$ and $\exp(\cdot)$ are not invoked as discussed in \cite{Robertson}. 
The quality of this approximation depends on how many values of $f_c(\cdot)$ are stored.
\textcolor{black}{In the numerical evaluation of Section VII}, we use 256 values for $f_c(x)$ uniformly spaced in the range $0\leq x \leq 10$, which is sufficient to provide a performance as good as the exact one.

\subsection{Max-Log-MAP Detector}
The Max-Log-MAP algorithm is a sub-optimal solution that takes the maximum argument of the exponential terms as $\ln(e^{\delta_0}+e^{\delta_1}+ \cdots e^{\delta_{N-1}}) \approx \max_i{\delta_i}$.
The Max-Log-MAP detector simply computes
\begin{equation}\label{eq:Le_3}
	\begin{split}
	L^{{\rm e,I}}_{p,q,b} =  \max_{\ma{z} \in \mathcal{R}^Q_{q,b,0}}\left(t^{\rm I}_{p}(\ma{z})\right) 
 - \max_{\ma{z} \in \mathcal{R}^Q_{q,b,1}}\left(t^{\rm I}_{p}(\ma{z})\right)-L^{{\rm a,I}}_{p,q,b}.
	\end{split}
\end{equation}
The differences with the Log-MAP detector are that the Max-Log-MAP method does not require the additions involved in the argument of $f_c(|\delta_0 - \delta_1|)$, and it does not require the look-up table operations. 

\section{\ac{SILE-EPIC} Detector}\label{sec:sile_epic}
In this section, the SILE-EPIC receiver presented in \cite{Sahin} is described.
Assuming \ac{i.i.d.} information bits, and focusing on the joint estimation of the data symbols $d_n$ and coded bits $c_{n,b}$, we can write the posterior probability factorization
\begin{equation}\label{eq:p_cdy}
	p(\ma{c},\ma{d}|\ma{y}) = p(\ma{y}|\ma{d})\prod_{n=0}^{N-1} p(d_n|\ma{c}_n) \prod_{b=0}^{K-1}p(c_{n,b}),
\end{equation}
from which the channel, mapping, and coding constraints are observed, which are respectively related to the terms $p(\ma{y}|\ma{d})$, $p(d_n|\ma{c}_n)$ and $\prod_{b=0}^{K-1}p(c_{n,b})$. 
In the context of a factor graph model, these constraints respectively define the factor nodes (FNs) {\it EQU Node}, {\it DEM Node}, and {\it DEC Node}.
As such, this iterative receiver works as a message-passing algorithm that operates on the variables $d_n$ and $c_{n,b}$ with the constraints determined by the FNs. 

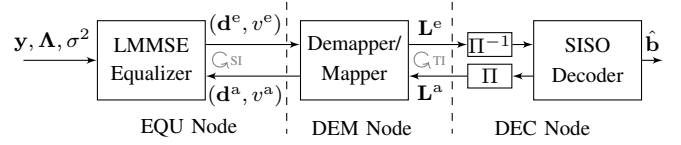
\begin{figure}[t!]
	\tikzstyle{block} = [draw, fill=white, rectangle, 
minimum height=3em, minimum width=3em]
\tikzstyle{block2} = [draw, fill=white, rectangle, 
minimum height=1em, minimum width=2.4em]

\tikzstyle{block3} = [draw, fill=white, rectangle, 
minimum height=0.9em, minimum width=0.2em]

\tikzstyle{multiplier} = [draw,circle,fill=blue!20,add={}{}{}{}] {} 
\tikzstyle{sum} = [draw,circle,fill=blue!20,add2={}{}{}{}] {} 
\tikzstyle{input} = [coordinate]
\tikzstyle{output} = [coordinate]
\tikzstyle{pinstyle} = [pin edge={to-,thin,black}]

\def\windup{
	\tikz[remember picture,overlay]{
		\draw (-0.8,0) -- (0.8,0);
		\draw (-0.8,-0.4)--(-0.4,-0.4) -- (0.4,0.4) --(0.8,0.4);
}}

\centering
\begin{tikzpicture}[auto, node distance=2cm,>=latex']
		
	\node [input, name=input2] at (0,0) {};
	
	\node [block, right of=input2,node distance=1.7cm] (mmse_eq) {\color{white}{\footnotesize SWH Det.}};
	\node [above of=mmse_eq,node distance=0cm,yshift=0.2cm] () {\footnotesize LMMSE};
	\node [above of=mmse_eq,node distance=0cm,yshift=-0.2cm] () {\footnotesize Equalizer};
	\node [below of=mmse_eq,node distance=0.9cm,xshift=0.5cm] () {\footnotesize EQU Node};
	
	\node [right of=mmse_eq, node distance=0.6cm,yshift=0.21cm] (mmse_eq_top_dummy) {};
	\node [right of=mmse_eq, node distance=0.6cm,yshift=-0.21cm] (mmse_eq_bottom_dummy) {};

	\node [block, right of=mmse_eq, node distance=2.7cm] (mapper_demapper) {\color{white}{\footnotesize SWH Det.}};
	\node [above of=mapper_demapper,node distance=0cm,yshift=0.2cm] () {\footnotesize Demapper/};
	\node [above of=mapper_demapper,node distance=0cm,yshift=-0.2cm] () {\footnotesize Mapper};
	\node [below of=mapper_demapper,node distance=0.9cm,xshift=0.1cm] () {\footnotesize DEM Node};	

	\node [left of=mapper_demapper, node distance=0.6cm,yshift=0.21cm] (mapper_demapper_dec_top_dummy) {};
	\node [left of=mapper_demapper, node distance=0.6cm,yshift=-0.21cm] (mapper_demapper_dec_bottom_dummy) {};
	
	\node [right of=mapper_demapper, node distance=0.6cm,yshift=0.21cm] (mapper_demapper_dec_top_dummy2) {};
	\node [right of=mapper_demapper, node distance=0.6cm,yshift=-0.21cm] (mapper_demapper_dec_bottom_dummy2) {};

	\node [block3, right of=mapper_demapper,node distance=1.8cm,yshift=0.21cm] (deinterleaver) {\color{white}{\footnotesize .....}};
	\node [right of=deinterleaver,node distance=0cm] () {\color{black}{\footnotesize $\Pi^{-1}$}};
		\node [block3, right of=mapper_demapper,node distance=1.8cm,yshift=-0.21cm] (interleaver) {\color{white}{\footnotesize .....}};
	\node [right of=interleaver,node distance=0cm] () {\color{black}{\footnotesize $\Pi$}};
	
	\node [block, right of=mapper_demapper, node distance=3.1cm] (siso_dec) {\color{white}{\footnotesize SWH Det.}};
	\node [above of=siso_dec,node distance=0cm,yshift=0.2cm] () {\footnotesize SISO};
	\node [above of=siso_dec,node distance=0cm,yshift=-0.2cm] () {\footnotesize Decoder};
	\node [below of=siso_dec,node distance=0.9cm,xshift=-0.6cm] () {\footnotesize DEC Node};	
	
	\node [left of=siso_dec, node distance=0.6cm,yshift=0.21cm] (siso_dec_top_dummy) {};
	\node [left of=siso_dec, node distance=0.6cm,yshift=-0.21cm] (siso_dec_bottom_dummy) {};
	
	\draw [->] (input2) -- node [xshift=-0.1cm]{\footnotesize $\mathbf{y}, \mathbf{\Lambda}, \sigma^2$} (mmse_eq);
	
	\draw [->] (mmse_eq_top_dummy) -- node [xshift=-0.1cm]{\footnotesize $(\ma{d}^{\rm e},v^{\rm e})$} (mapper_demapper_dec_top_dummy);
	\draw [->] (mapper_demapper_dec_bottom_dummy) -- node [xshift=-0.1cm,yshift=0.0cm]{\footnotesize $(\ma{d}^{\rm a},v^{\rm a})$} (mmse_eq_bottom_dummy);
	
	\draw [->] (mapper_demapper_dec_top_dummy2) -- node [xshift=-0.1cm]{\footnotesize $\ma{L}^{\rm e}$} (deinterleaver);
	\draw [->] (interleaver) -- node [xshift=-0.1cm]{\footnotesize $\ma{L}^{\rm a}$} (mapper_demapper_dec_bottom_dummy2);
	
	\draw [->] (deinterleaver) -- node [xshift=-0.1cm]{} (siso_dec_top_dummy);
	\draw [->] (siso_dec_bottom_dummy) -- node [xshift=-0.1cm]{} (interleaver);	
	
	\node [output, name=output2, right of=siso_dec,node distance=1cm] {};
	\draw [->] (siso_dec) -- node [xshift=0cm]{\footnotesize $\hat{\mathbf{b}}$} (output2);

	\node [left of=mapper_demapper, node distance=0.9cm,yshift=0.9cm] (eq_box_top) {};
	\node [left of=mapper_demapper, node distance=0.9cm,yshift=-1.1cm] (eq_box_bottom) {};
	\draw [-,dashed] (eq_box_top) -- node []{} (eq_box_bottom);	
	
	\node [right of=mapper_demapper, node distance=1.3cm,yshift=0.9cm] (dem_box_top) {};
	\node [right of=mapper_demapper, node distance=1.3cm,yshift=-1.1cm] (dem_box_bottom) {};
	\draw [-,dashed] (dem_box_top) -- node []{} (dem_box_bottom);

	   \draw[-,gray] ($(-0.15+3,0+0.03)+(-0.15+0,0+0.03)$) arc
	[
	start angle=30,
	end angle=330,
	x radius=0.1cm,
	y radius =0.1cm
	] ;
	
	\draw[-to,gray] ($(-0.15+3.007,0+0.03)+(-0.15+0,-0.088+0.03)$) arc
	[
	start angle=330,
	end angle=331,
	x radius=0.1cm,
	y radius =0.1cm
	] ;
	
	\node [right of=mmse_eq,node distance=1.13cm,xshift=0.0cm,gray] () {\tiny SI};

		   \draw[-,gray] ($(3+1.19,0+0.03)+(0+1.19,0+0.03)$) arc
	[
	start angle=30,
	end angle=330,
	x radius=0.1cm,
	y radius =0.1cm
	] ;
	
	\draw[-to,gray] ($(3.007+1.19,0+0.03)+(0+1.19,-0.088+0.03)$) arc
	[
	start angle=330,
	end angle=331,
	x radius=0.1cm,
	y radius =0.1cm
	] ;
	
		\node [right of=mapper_demapper,node distance=1.13cm,xshift=0.0cm,gray] () {\tiny TI};
	
\end{tikzpicture}
	\vspace{-0.9cm} 
	\caption{Block diagram of \ac{SILE-EPIC} receiver of \cite{Sahin}.}
	\label{fig:sile_epic_detector}
\end{figure}

In the following, we directly describe the messages derived in \cite{Sahin} between the factor and variable nodes. A diagram of the \ac{SILE-EPIC} detector is depicted in Fig.~\ref{fig:sile_epic_detector}.

\subsection{Messages for DFT Precoder}\label{subsec:messages_DFT}
\subsubsection{Messages from DEC to DEM}
As discussed in the Subsection \ref{subsec:iterative_receiver}, the DEC node is a SISO decoder that feeds back a-priori LLRs $\ma{L}^{{\rm a}}$ to the DEM.
Considering the mapping constraint, the prior \ac{PMF} of $d_n$ is computed as
\begin{equation}\label{eq:Pa}
	P_n({\rm d}) \propto \prod_{b=0}^{\log_2 \! J -1}  \exp\left(-\phi_b^{-1}({\rm d}) L^{{\rm a}}_{n,b}\right), \,\, \forall {\rm d}\in \mathcal{D}.
\end{equation}
For completeness, the indexes $n$ and $b$ of $L^{{\rm a,I}}_{n,b}$ correspond to the $n$th QAM symbol and $b$th bit.
\subsubsection{Messages from DEM to EQU}
This subsection describes how the variables $(\ma{d}^{\rm a},v^{\rm a})$ in Fig.~\ref{fig:sile_epic_detector} are updated, given that the DEM has the PMF in \eqref{eq:Pa} and the variables $(\ma{x}^{\rm e},v^{\rm e})$ as inputs.
Firstly, the DEM node computes the posterior symbol PMF as
\begin{equation}\label{eq:Dn}
	D_n({\rm d}) \propto \exp(|{\rm d} - d^{\rm e}_n|^2/v^{\rm e})P_n({\rm d}), \,\, \forall {\rm d}\in \mathcal{D},
\end{equation}
from which the moments are computed as
\begin{equation}\label{eq:mu_n}
\mu^{\rm a}_n = \sum_{{\rm d}\in \mathcal{D}} {\rm d}D_n({\rm d}),
\end{equation}
and
\begin{equation}\label{eq:gamma_a_n}
	\gamma^{\rm a}_n = \sum_{{\rm d}\in \mathcal{D}} |{\rm d}|^2D_n({\rm d}) - |\mu^{\rm a}_n|^2.
\end{equation}
Since the receiver assumes that the variable nodes related to $\ma{d}$ are multivariate white Gaussian random variables, the second moment of all elements of $\ma{d}$ are approximated to the mean
\begin{equation}\label{eq:gamma_a}
	\bar{\gamma}^{\rm a} = \frac{1}{N}\sum_{n=0}^{N-1}\gamma^{\rm a}_n.
\end{equation}
The variable pair $({d}_n^{\star},v^{\star})$ is derived by dividing the PDF $\mathcal{CN}({\mu}_n^{\rm a},\gamma^{\rm a})$ by $\mathcal{CN}({d}_n^{\rm e},\gamma^{\rm e})$, which results in
\begin{equation}\label{eq:xstar}
	{d}^{\star}_n = \frac{{\mu}^{\rm a}_nv^{\rm e} - {d}^{\rm e}_n\bar{\gamma}^{\rm a}}{v^{\rm e} - \bar{\gamma}^{\rm a}}, \,\,\, \text{and} \,\,\, v^{\star} = \frac{v^{\rm e}\bar{\gamma}^{\rm a}}{v^{\rm e} - \bar{\gamma}^{\rm a}}.
\end{equation}
The scheme above can lead to negative variance when $\bar{\gamma}^{\rm a}>v^{\rm e}$. In this case, we use the values $({d}^{\rm a}_n,v^{\rm a})$ as \cite{Sahin}.

As highlighted in \cite{Sahin}, using $({d}^{\star}_n,v^{\star})$ directly as feedback leads to detrimental local extrema. 
To circumvent this problem, the linear smoothing procedure below is used
\begin{equation}\label{eq:xa_va}
	\begin{split}
		& {d}^{\rm a(next)}_n = (1-\beta){d}^{\star}_n + \beta {d}^{\rm a(prev)}_n \\
		& v^{\rm a(next)} = (1-\beta)v^{\star} + \beta v^{\rm a(prev)},
	\end{split}
\end{equation}	
for $0 \leq \beta \leq 1$. We set $\ma{d}^{\rm a(prev)} = \ma{0}_N$ and $v^{\rm a(prev)} = 1$ at the beginning of each new self-iteration between DEM and EQU.

\subsubsection{Messages from EQU to DEM}
This subsection describes how the EQU node updated the variables $(\ma{d}^{\rm e},v^{\rm e})$, given that it has $(\ma{d}^{\rm a},v^{\rm a})$ as input from the DEM node.
The equalization process of the EP based receiver turns out to be equivalent to the \ac{CWCU} LMMSE \cite{Huemer,Bomfin_WCNC2}, which is an unbiased estimator, where the symbol mean and error variance are computed as
\begin{equation}\label{eq:x_e}
	\begin{split}
		{\ma{d}}^{\rm e} \! = {\ma{d}}^{\rm a} +
		\frac{1}{\lambda}\mathbf{F}^{\rm H}_N{\mathbf{\Lambda}}^{\rm H}\left({\mathbf{\Lambda}}{\mathbf{\Lambda}}^{\rm H}v^{\rm a} \! +\! \sigma^2\mathbf{I}_N\right)^{-1}\left(\mathbf{y} \!- \!{\mathbf{\Lambda}}\mathbf{F}_N{\ma{d}}^{\rm a}\!\right )
	\end{split}
\end{equation}
and
%
\begin{equation}\label{eq:v_e}
	\begin{split}
				v^{\rm e} \!   =\frac{1}{\lambda} -  v^{\rm a},
	\end{split}
\end{equation}
where the normalization variable
\begin{equation}\label{eq:lambda}
\lambda = \frac{1}{N}{\rm Tr}({\mathbf{\Lambda}}^{\rm H}\left( {\mathbf{\Lambda}}{\mathbf{\Lambda}}^{\rm H}v^{\rm a} + \!\sigma^2\mathbf{I}_N\right )^{-1}\!{\mathbf{\Lambda}}),
\end{equation}
can be easily computed as shown in \cite{Bomfin_WCNC2}.

\subsubsection{Messages from DEM to DEC}
Lastly, the DEM node computes the extrinsic LLRs as input to the DEC node as
\begin{equation}\label{eq:Le_sile_epic}
	L^{{\rm e}}_{n,b} = \ln{\frac{\sum_{{\rm d} \in \mathcal{D}_{b}^{(0)}} D_n({\rm d}) }{\sum_{{{\rm d}} \in \mathcal{D}_{b}^{(1)}} D_n({\rm d})}}-L^{{\rm a}}_{n,b}.
\end{equation}

\subsection{Messages for the Sparse Precoders}
The extension of the SILE-EPIC to the sparse precoders is fairly straightforward. The exchanged messages between the EQU and DEM nodes are now derived for the model of \eqref{eq:yp}, which groups the symbols that are spread over the same set of sub-carriers. 
The messages exchanged between the DEM and DEC nodes remain unchanged since they do not directly depend on the modulation matrix.

\subsubsection{Messages from DEM to EQU}
The equation \eqref{eq:gamma_a} is modified as
\begin{equation}\label{eq:gamma_a_p}
	\bar{\gamma}^{\rm a}_p = \frac{1}{Q}\sum_{n\in I_p}\gamma^{\rm a}_{n},
\end{equation}	
where the error variance is averaged only for indexes whose symbols share the same sub-carriers.
Next, using the same subvectors\footnote{The $p$th subvector is related to the symbols that are spread over the same set of subcarriers.} definition of Section \ref{sec:sparse_precoders}, we compute $\ma{d}^{\star}_p$ and $v^{\star}_p$ analogously to \eqref{eq:xstar}, where we now use $\boldsymbol{\mu}^{\rm a}_p$, $\ma{d}^{\rm e}_p$, $v^{\rm e}_p$ and $\bar{\gamma}^{\rm a}_p$.
Then, $\ma{d}^{\star}_p$, $v^{\star}_p$, $\ma{d}^{\rm e}_p$ and $v^{\rm e}_p$ are applied analogously to the equation \eqref{eq:xa_va} to compute $\ma{d}^{\rm a}_p$ and $v^{\rm a}_p$. 

\subsubsection{Messages from EQU to DEM}
The message from the EQU to DEM $(\ma{x}^{\rm e}_p$, $v^{\rm e}_p)$ are computed with equations \eqref{eq:x_e}, \eqref{eq:v_e} and \eqref{eq:lambda} by replacing  $\ma{d}^{\rm a}$, $v^{\rm a}$, $\ma{F}_N$, $\ma{I}_N$, $\ma{\Lambda}$ and $N$ by $\ma{x}^{\rm a}_p$, $v^{\rm a}_p$, $\ma{F}_Q$, $\ma{I}_Q$, $\ma{\Lambda}_p$ and $Q$, respectively.
Obviously, for the SWH precoder, $\ma{W}_Q$ is used instead of $\ma{F}_Q$.

\subsection{Scheduling}
There are two iteration loops in the structure of the SILE-EPIC receiver, see Fig.~\ref{fig:sile_epic_detector}. 
In particular, there is the self-iteration (SI) loop, which represents the messages exchanged by the EQU and DEM nodes, and there is the turbo iteration (TI) loop, which is related to the messages exchanged between the DEM and DEC nodes.  
Scheduling refers to the order in which these loops are performed.
Defining the $N_{\tau}$ and $N_{\rm s}$ as the maximum number of turbo and self-iteration, the indexes are conventionally defined as $\tau \in \{0,1,\cdots,N_{\tau}\}$ and $s \in \{0,1,\cdots,N_{\rm s}\}$ to represent the turbo and self-iterations indexes, respectively. 
Thus, there are $N_{\rm s}+1$ SIs for each TI index, which is determined empirically for a targeted performance and complexity trade-off.

\subsection{Comments on Full vs Sparse Spreading}\label{subsec:full_vs_sparse_spreading}
In \cite{BomfinSWH}, the sparse spreading has been explored from the complexity perspective, where it has been shown that the equalizer's complexity is reduced from $\mathcal{O}(N \log N)$ to $\mathcal{O}(N \log Q)$, which can be considerable since typically we can set $Q\ll N$.
Another aspect that we investigate in this work is the performance gain due to a more accurate approximation of $\gamma^{\rm a}_p$ than $\gamma^{\rm a}$.	
As shown in Appendix \ref{appendix:variance_approx}, when the spreading precoding scheme is used, $\gamma^{\rm a}$ in \eqref{eq:gamma_a} is averaged over the variance of all $N$ data symbols, which is a worse approximation than the sparse spreading one,  $\gamma^{\rm a}_p$ in \eqref{eq:gamma_a_p}, which averages over the subset of $Q$ symbols.

\begin{table*}
	\centering
	\small
	\footnotesize
	\caption{Number of real multiplications and additions per QAM symbol.}
	\vspace{-3mm}
	\begin{tabular}{rcc}\toprule
		Precoder and Receiver	& Additions & Multiplications  \\\midrule
		SWH-Log-MAP  & $(3\log_2 J + 2) J^{\frac{Q}{2}} + 2 (Q (\sqrt{J} - 1) + 1))$ &  $6(Q(\sqrt{J}-1)+1)$  \\
		SWH-Max-Log-MAP & $  (\log_2 J + 2) J^{\frac{Q}{2}} + 2 (Q (\sqrt{J} - 1) + 1))$ & $4(Q(\sqrt{J}-1)+1)$ \\
		SWH-SILE-EPIC  & $(N_{\rm s}+1)(4 \log_2 Q + \log_2 J + 7J + 6)$ &  $(N_{\rm s}+1)(8J + 11)$ \\ 
		DFT-SILE-EPIC  & $(N_{\rm s}+1)(4 \log_2 Q + \log_2 J + 7J + 6)$ &  $(N_{\rm s}+1)(4\log_2 Q + 8J + 11)$ \\
		SDFT-SILE-EPIC  & $(N_{\rm s}+1)(4 \log_2 N + \log_2 J + 7J + 6)$ & $(N_{\rm s}+1)(4\log_2 N + 8J + 11)$  \\ \bottomrule
	\end{tabular}\label{tab:complexity_analysis}
\end{table*}

\section{Complexity Analysis}\label{sec:complexity_analysis}
In the following, the complexity of the MAP and SILE-EPIC receivers respectively presented in Sections \ref{sec:map_detectors_SWH} and \ref{sec:sile_epic} are derived in terms of real additions and multiplications per QAM symbol. 
The corresponding values are summarized in Table \ref{tab:complexity_analysis}. 
\textcolor{black}{Also, the different methods are compared in terms of complexity in Section VII.}


%
\subsection{Log-MAP for SWH}\label{subsec:log_map_complexity}
Starting with the number of additions, the computation of $C^{\rm I}_{p}(q',i)$ in \eqref{eq:C} requires $(Q(\sqrt{J}-1)+1) Q$ subtractions for all $q' \in \{0,1,\cdots, Q-1\}$ and $i \in \{0,1,\cdots  (Q(\sqrt{J}-1)+1) \} $, that is computed for a group of $Q$ symbols. 
Also, there are $Q J^{\frac{Q}{2}}$ additions to compute the sum over $C^{\rm I}_{p}(q',i)$ in \eqref{eq:t_pq} for all $\ma{z} \in \mathcal{R}^Q$.
The incorporated sum of a-priori LLRs in \eqref{eq:t_pq} results in  $Q J^{\frac{Q}{2}} \frac{1}{2}\log_2{{J}} $ additions for all $\ma{z} \in \mathcal{R}^Q$.
Finally, since each symbol has $\frac{1}{2}\log_2 J$ bits, there are $2 Q J^{\frac{Q}{2}} \frac{1}{2}\log_2 J $ additions for each group of $Q$ symbols to compute the LogSumExp using \eqref{eq:log_sum}, as described in Subsection~\ref{subsec:logMAP_det}.
Thus, the total amount of additions is found as $ 2 (\frac{3}{2}\log_2 J + 1) J^{\frac{Q}{2}} + 2 (Q (\sqrt{J} - 1) + 1))$, where the above terms have been added, multiplied by two to account for the in-phase and quadrature components, and divided by $Q$ to have the quantity per symbol.

Regarding the multiplications, we see that $C_{p,q'}(s_i)$ in \eqref{eq:C} requires $3(Q(\sqrt{J}-1)+1) Q$ multiplications for all $q'$ and $i$, where the number 3 is multiplied to account for 
the element-wise products related to the channel, square exponent, and noise variance division.
The final value of $6(Q(\sqrt{J}-1)+1)$ is obtained by multiplying the above quantity by two to account for the in-phase and quadrature components, and dividing it by $Q$ to have the quantity per symbol.

Although the overall complexity is considerably reduced due to the small number of multiplications, there are two important remarks to be made.
Firstly, it is clear that this solution does not decrease the number of additions necessary to marginalize the posterior probability which scales with $J^{\frac{Q}{2}}$.
Secondly, the receiver still needs to store the sets $\mathcal{S}(\ma{z},q)$ in a memory for each $\ma{z}$. Since there are $J^{\frac{Q}{2}}$ possible realizations of $\ma{z}$, the number of values to be stored can be prohibitively high which also prevents the scalability of this detector.
For the above reasons, the Log-MAP receiver for SWH is restricted to small values of $Q$ and $J$, as shown in Section~\ref{sec:numerical_evaluation}.
Thus, it is clear that in order to expand the range of $Q$ and $J$ in which the Log-MAP can be implemented, it is necessary to mitigate these issues.
In particular, the number of additions could be decreased by allowing only the values of $\ma{z}$ which have a significant impact on the LogSumExp.
Regarding the memory sets $\mathcal{S}(\ma{z},q)$, a possible approach to decrease the total amount of saved data could involve symmetries in the values of $\ma{z}$. 
A clear example is that for each $\ma{z}$ there is a negative $\ma{z}' = -\ma{z}$ which the index $i'\in\mathcal{S}(\ma{z}',q)$ can be determined from $\mathcal{S}(\ma{z},q)$.
This simple approach decreases by half the amount of necessary values $\mathcal{S}(\ma{z},q)$ to be saved.

\subsection{Max-Log-MAP for SWH}
For the Max-Log-Max method, a very similar approach to the previous subsection is taken, therefore we just highlight the differences with the Log-MAP.
Regarding the number of additions, the Max-Log-MAP does not require the operations involved in the argument of $f_c(|\delta_0 - \delta_1|)$, thus it has $2 J^{\frac{Q}{2}} \log_2 J $ less additions per QAM symbol.
For the number of multiplications, the values $C_{p,q'}(s_i)$ in \eqref{eq:C} can be computed with $2(Q(\sqrt{J}-1)+1)$ multiplications where the number 2 is multiplied to account only for the element-wise products related to the channel and square exponent because the noise variance division can be done after the $\max(\cdot)$ operation in \eqref{eq:Le_3}.

For the same reasons discussed in the previous subsection, the Max-Log-MAP receiver is also restricted to small values of $Q$ and $J$. Thus, future research should be made to allow a more diverse range of $Q$ and $J$.

\subsection{SWH-SILE-EPIC}
%
Firstly, we note that the a-priori symbol probability $P_n({\rm d})$ in equation \eqref{eq:Pa} can be integrated into the computation of $D_n({\rm d})$ in \eqref{eq:Dn}.
Thus, \eqref{eq:Dn} needs $\log_2 J$ due to \eqref{eq:Pa}, $2 J$ due to the argument of \eqref{eq:Dn}\footnote{Additions (or subtractions) between two complex numbers count twice.} for all ${\rm d}\in \mathcal{S}$ and $J$ to normalize \eqref{eq:Dn}.
Equations \eqref{eq:mu_n}, \eqref{eq:gamma_a_n}, \eqref{eq:gamma_a}, \eqref{eq:xstar} and \eqref{eq:xa_va} requires, $2J$, $2J+1$, $1$, $2$ and $2$, respectively.
Adding the above components and  multiplying the number of self iterations $N_{\rm s}$, we find that $N_{\rm s}(\log_2 J + 7J + 6)$ additions per symbol to compute the message from the DEM to EQU $(\ma{x}^{\rm e},v^{\rm e})$. 
In order to compute the operations spent in \eqref{eq:x_e}, only the complexity of the modulation $\ma{W}_Q$ is considered for simplicity.
We note that $\ma{W}_Q$ requires $2 N \log_2 Q$ real addition \cite{BomfinSWH}.
Since it is called two times, by dividing it by $N$, we find $2 \log_2 Q$ additions for the precoding.
Since the above operations are computed for each iteration index $s \in \{0,1,\cdots,N_{\rm s}\}$, they are processed $(N_{\rm s}+1)$ number of times in total.
Thus, from the equation \eqref{eq:Dn} to \eqref{eq:x_e}, there are $(N_{\rm s}+1)(4 \log_2 Q + \log_2 J + 7J + 6)$ real additions.

Regarding the amount of multiplications, equation \eqref{eq:Dn} require $2J\,$\footnote{The square of a complex counts two multiplications.} operations due to the square $(\cdot)^2$, and $J$  $v^{\rm e}$  for all ${\rm d}\in \mathcal{S}$ and $J$ to normalize \eqref{eq:Dn}. 
The remaining equations \eqref{eq:mu_n}, \eqref{eq:gamma_a_n}, \eqref{eq:xstar} and \eqref{eq:xa_va} require $2J\,$\footnote{Multiplications (or divisions) between real and complex numbers count twice.}, $2J +1$, $6$ and $4$, respectively.
Adding the above components and  multiplying the number of self-iterations $N_{\rm s}$, we find that $N_{\rm s}(6J + 11)$ additions per symbol to compute the message from the DEM to EQU.
Regarding the equalizer in \eqref{eq:x_e}, $\ma{W}_Q$ spends no multiplication, thus the overall multiplication amount is $(N_{\rm s}+1)(8J + 11)$.

\subsection{SDFT-SILE-EPIC}
The amount of additions spent by the SDFT-SILE-EPIC is exactly the same as SWH-SILE-EPIC.
Basically, the number of additions and multiplications spent from equation \eqref{eq:Dn} to \eqref{eq:xa_va} is the same regardless of the precoder.
And the equalizer in \eqref{eq:x_e}, the amount of additions is also the same as SWH because the transformation is done with the same size.
Thus, the SDFT-SILE-EPIC consumes $(N_{\rm s}+1)(4 \log_2 Q + \log_2 J + 7J + 6)$ real additions.

For the multiplications, the DFT consumes $2 N \log_2 Q$ real multiplications\footnote{The $Q$-size FFT consumes $1/2 Q \log_2 Q$ complex multiplications, which is multiplied by 4 to account for real multiplications.}. Since it is called twice in \eqref{eq:x_e}, there are $4 \log_2 Q$ more multiplications in the SDFT than in the SWH per QAM symbol, which is $(N_{\rm s}+1)(4\log_2 Q + 6J + 11)$.

\subsection{DFT-SILE-EPIC}
Again, what changed from SDFT to DFT, is the transformation $\ma{F}_N$ in \eqref{eq:x_e} instead of $\ma{F}_Q$.
Thus, comparing with the results valued for SDFT-SILE-EPIC, it is straightforward to derive the number of additions and multiplications for the DFT-SILE-EPIC respectively as $(N_{\rm s}+1)(4 \log_2 N + \log_2 J + 7J + 6)$ and $(N_{\rm s}+1)(4\log_2 N + 6J + 11)$.

\section{Numerical Evaluation}\label{sec:numerical_evaluation}

Simulation parameters are described in Table~\ref{tab:simulation_parameters}. 
The amount of QAM symbols per transmission is $N=256$.
The channel is the highly frequency selective Proakis-C impulse response $h = [0.23 \,\, 0.46 \,\, 0.69 \,\, 0.46 \,\, 0.23]^{\rm T}$ with maximum delay of $L=5$ samples.
In the frequency domain, the channel is ${\rm diag}(\ma{\Lambda}) =  {\rm DFT}_N (h)$ with zero on its off-diagonals.
We set $Q=\{4,8\}$ for the sparse spreading precoding schemes. 
Notice that the condition $Q\geq L$ \cite{BomfinSWH} is respected only for $Q=8$.
Regarding the encoder and decoder, the $R=1/2$ code rate recursive systematic convolutional (RSC) code is employed with polynomial $[1,5/7]_8$, and the SISO decoder is used with BCJR algorithm.
The number of turbo iterations is set $N_{\tau} = 9$ for all cases.

The systems are evaluated in terms of \ac{FER} and \ac{EXIT} chart.
\textcolor{black}{The \ac{EXIT} chart analyses the asymptotic behavior of the system \cite{Brink}}, which is meaningful under the assumptions of long codeword and a high number of turbo iterations.
\textcolor{black}{In addition, the EXIT chart provides an assessment of the equalizer independently of the encoder and decoder, making the conclusions more general.}
The detector's asymptotic behavior is characterized by the transfer curve $I_{\rm E,det} = \mathcal{T}_{\rm det}(I_{\rm A,det})$, where $I_{\rm E,det}$ and $I_{\rm A,det}$ are the \ac{MI} between the coded data with the a-priori and extrinsic LLRs, respectively.
A typical approach to compute $\mathcal{T}_{\rm det}(I_{\rm A,det})$ and $I_{\rm E,dec} = \mathcal{T}_{\rm dec}(I_{\rm A,dec})$ is to assume Gaussian distributed LLRs in the input of the respective blocks, i.e., detector and decoder. 
However, in practice the LLRs' distributions differ from the Gaussian assumption, so the convergence of iterative receivers can be analyzed with the \ac{MI} trajectories w.r.t. the turbo iterations using finite length simulations \cite{EXIT2}, where the inverse transfer curve of the decoder $\mathcal{T}^{-1}_{\rm dec}(I_{\rm A,det})$ is also plotted, \textcolor{black}{which are computed in relatively high SNR such that we are able to observe the receiver's convergence.}

\begin{table}[t!]
	\centering
	\footnotesize
	\caption{Simulation Parameters.}
	\vspace{-3mm}	
	\begin{tabular}{c|c}\toprule
		Parameter & Value  \\ \midrule
		number of symbols per block & $N=256$ \\
		channel & Proakis-C, $L=5$ \\
		spreading parameter & $Q=\{4,8\}$ \\
		QAM order & $J=\{4,16,64\}$ \\
		code rate & $R=1/2$  \\
		encoder & RSC $[1,5/7]_8$ \\
		decoder & SISO BCJR (log-MAP)  \\ \bottomrule
	\end{tabular}\label{tab:simulation_parameters}
\end{table}

\begin{table}[t!]
	\centering
	\footnotesize
	\caption{Number of Real Additions per QAM symbol.}
	\vspace{-3mm}	
	\begin{tabular}{lr|c|c|cc}\toprule
		Precoder and Receiver 				& 	& {QPSK}	& {16-QAM} 	& 64-QAM \\ \midrule
		SWH & $Q = 4$ 					& 138  		&  	3,610	&  81,978 \\
		Log-MAP			& $Q = 8$		& 	2,066	&  	917,554	&  \textcolor{black}{335,544,434} 	 \\ \midrule
		SWH & $Q = 4$ 					&  	74		&  	1,562	&  32,826 \\
		Max-Log-MAP			& $Q = 8$ 	& 	1,042	&  	393,266	&  134,217,842 	 \\ \midrule
		SWH-SILE-EPIC  &  $Q=8$ 		&  	144		&  	804		&  3,304	 \\ \midrule
		SDFT-SILE-EPIC &  $Q=8$ 		&  	144		&  	804		&  3,304	 \\ \midrule
		DFT-SILE-EPIC  &     			&  	204		&  	924		&  3,444	 \\ \bottomrule
	\end{tabular}\label{tab:real_add}
\end{table}

\begin{table}[t!]
	\centering
	\footnotesize
	\caption{Number of Real Multiplications per QAM symbol.}
	\vspace{-3mm}	
	\begin{tabular}{lr|c|c|cc}\toprule
		Precoder and Receiver 				& 	& {QPSK}	& {16-QAM} 	& 64-QAM \\ \midrule
		SWH & $Q = 4$ 					&   30		&  	78		&   174	 \\
		Log-MAP			& $Q = 8$		& 	54		&  	150		&  	342 	 \\ \midrule
		SWH & $Q = 4$ 					&  	20		&  	52		&   116	 \\
		Max-Log-MAP			& $Q = 8$ 	& 	36		&  	100		&  	228 	 \\ \midrule
		SWH-SILE-EPIC  &  $Q=8$ 		&  	129		&  	834		&   3,661 \\ \midrule
		SDFT-SILE-EPIC &  $Q=8$ 		&  	165		&  	906		&   3,745 \\ \midrule
		DFT-SILE-EPIC  &     			&  	225		&  	1,026	&   3,885 \\ \bottomrule
	\end{tabular}\label{tab:real_multi}
\end{table}

\subsection{Evaluation of MAP receivers for SWH}
\begin{figure}[t!]
	\centering
	\includegraphics[scale=0.9]{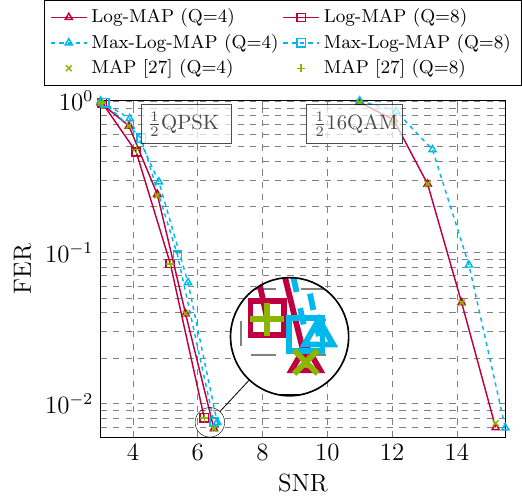}
	\vspace{-0.3cm}
	\caption{\textcolor{black}{Comparison between Log-MAP and Max-Log-MAP methods of the SWH detector described in Section \ref{sec:map_detectors_SWH}.}}
	\label{fig:FER_SWH_MAP}
\end{figure}

\begin{figure}[t!]
	\includegraphics[scale=0.9]{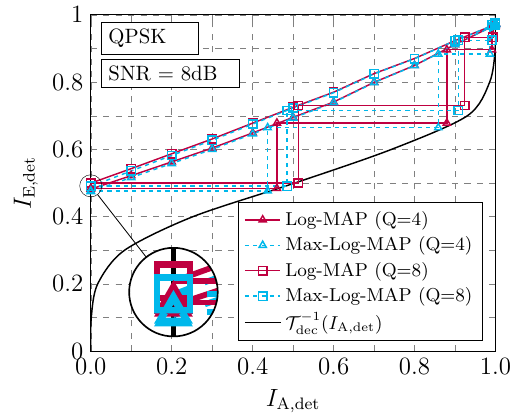}
	\includegraphics[scale=0.9]{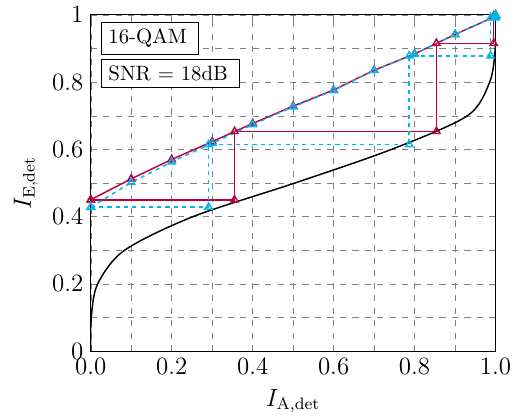}
	\vspace{-0.3cm}
	\caption{\textcolor{black}{Comparison between Log-MAP and Max-Log-MAP methods of the SWH detector described in Section \ref{sec:map_detectors_SWH}.}}
	\label{fig:EXIT_SWH_MAP}
\end{figure}

The MAP detectors for the SWH precoder shown in Section \ref{sec:map_detectors_SWH} are compared in terms of FER in Fig.~\ref{fig:FER_SWH_MAP}.
Firstly, we note that there is no performance loss in using the Log-MAP approximation in \ref{subsec:logMAP_det} compared to the general receiver of \cite{BomfinMAP}.
This result is expected since the Log-MAP approximation using the look-up table to compute $f_c(|\delta_0 - \delta_1|)$ is known to be accurate as shown in \cite{Robertson}. 
Regarding the Max-Log-MAP approximation, its implementation incurs a slight performance loss in relation to the Log-MAP of approximately 0.3 dB for QPSK with $Q=8$ and 16-QAM $Q=4$ for FER $=10^{-2}$. The gap is decreased for QPSK with $Q=4$.
These results indicate that the gap between the Log-MAP and Max-Log-MAP increases for higher $Q$ or $J$, which happens because the number of ignored $t^{\rm I}_{p}(\ma{z})$ values in \eqref{eq:Le2} increases.
In addition, examining the EXIT chart depicted in Fig.~\ref{fig:EXIT_SWH_MAP}, the Max-Log-MAP performance loss occurs due to the first iteration in all cases.
\textcolor{black}{In order to make this analysis more clear for the QPSK modulation, we magnified the results for $I_{\rm E,det}=\mathcal{T}_{\rm det}(I_{\rm A,det})$ for $I_{\rm A,det}=0$, where one can observe that $I_{\rm E,det}$ is higher for the Log-MAP method (red straight line).
For the 16-QAM results, the difference is more prominent.}
Basically, the Log-MAP equalizer feeds the decoder with LLRs that have higher MI with the coded data, leading to faster convergence in correctly detecting the transmitted codeword, which can be verified by the MI trajectories. 

The number of real additions and multiplications are shown in Tables \ref{tab:real_add} and \ref{tab:real_multi}, respectively.
Comparing the Max-Log-MAP against Log-MAP, we observe that the number of additions is decreased by approximately half in all cases, and the number of multiplications is decreased exactly by 1/3, which is clear from Table \ref{tab:complexity_analysis}. 
Since the performance is not decreased considerably, it makes the Max-Log-MAP method appealing in practical systems.
For the sake of completeness, we have also included the complexity for $J=64$. 
While the number of multiplications remains low, the number of additions increases exponentially with $J^{\frac{Q}{2}}$ and becomes impractical. 
As discussed in \cite{BomfinMAP}, this highlights the limitations of the MAP receiver when $Q$ or $J$ increases.
For the current approaches presented in this work, both Log-MAP and Max-Log-MAP methods are feasible for QPSK with $Q=4$ and 8, and 16-QAM with $Q=8$.
The bottleneck that limits the receiver is discussed in detail in Subsection \ref{subsec:log_map_complexity}, where directions to tackle this issue are provided for future works.

\begin{figure}[t!]
	\includegraphics[scale=0.9]{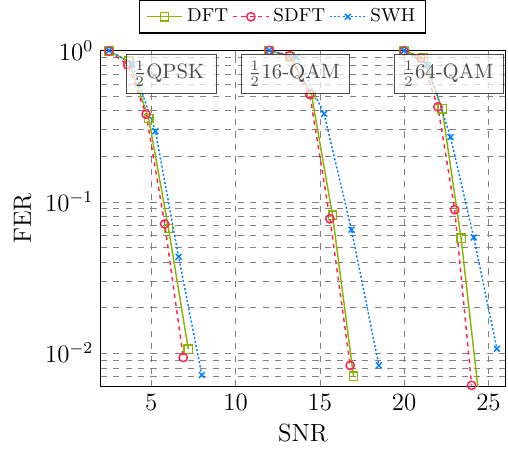}
	\includegraphics[scale=0.9]{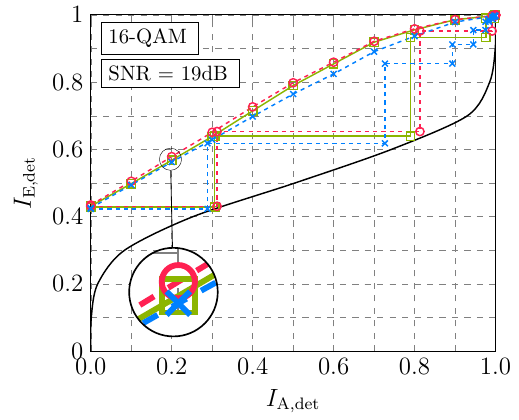}
	\vspace{-0.5cm}
	\caption{FER and EXIT curves for (S)DFT and SWH precoders with the SILE-EPIC receiver of \cite{Sahin} described in \ref{sec:sile_epic}. QPSK, 16-QAM and 64-QAM have respectively $N_{\rm s} = 2$, 5 and 6 self-iterations (SIs).}
	\label{fig:FER_SILE_EPIC}
\end{figure}

\subsection{Evaluation of Systems Employing SILE-EPIC}
This subsection investigates two different aspects, namely, i) the impact of sparsity using the DFT precoding, and ii) the comparison between the DFT vs WHT precoders with sparsity for $Q=8$.
We set $\beta$ in \eqref{eq:xa_va} as follows. 
Defining $N_{\tau}$ and $N_{\rm s}$ as the maximum number of turbo and self-iteration, the indexes $\tau \in \{0,1,\cdots,N_{\tau}\}$ and $s \in \{0,1,\cdots,N_{\rm s}\}$ represent the turbo and self-iterations indexes, respectively. 
In this work, our goal is not to optimize the best $\beta$ values, therefore we use the same formula as \cite{Sahin} which depends on the iteration indexes $(\tau,s)$.
In particular, we use $\beta_{\tau,s} = 0.7\cdot 0.9^{\tau + s}$ for QPSK, $\beta_{\tau,s} = 0.85\cdot 0.85^{\tau + s}$ for 16-QAM and $\beta_{\tau,s} = 0.85^{\tau + s}$ for 64-QAM.
Lastly, the QPSK, 16-QAM, and 64-QAM modulations have respectively $N_{\rm s} = 2$, 5 and 6 self-iterations (SIs).

Firstly, it is observed in Fig.~\ref{fig:FER_SILE_EPIC} that the SDFT system slightly outperforms the conventional DFT system. 
As explained in Subsection \ref{subsec:full_vs_sparse_spreading}, the system with sparsity has a better approximation to the a-priori variance of the data symbols.
This leads to a slight, but noticeable, improvement of approximately 0.3 dB at FER of $10^{-2}$ for the systems with QPSK and 64-QAM, and $0.1$ dB for the 16-QAM.
The results are also verified in the EXIT chart for 16-QAM, where we see that $\mathcal{T}_{\rm det}(0)$ is slightly higher for the SDFT system, which is sufficient to provoke faster receiver convergence, resulting in not only a better performance, but also decreasing the number of necessary turbo iterations.
Although it is not shown here, we have observed that this behavior is maintained for many other configurations of $\beta_{\tau,s}$, indicating that this is a general behavior.

Secondly, it is observed in Fig.~\ref{fig:FER_SILE_EPIC} that the SWH system performs worse than the SDFT scheme when both use the SILE-EPIC receiver. In particular, for the FER of $10^{-2}$, 1dB loss is observed for QPSK and 2dB loss occurs for 16-QAM and 64-QAM. 
Additionally, the EXIT chart also depicts the lack of convergence of the SWH system in comparison to the DFT systems, which is complementary to the FER curves.
This result has not been observed in previous works \cite{BomfinSWH}, where both SWH and DFT waveforms performed equally. 
Thus, the outcomes presented in this paper suggest that for highly selective channels such as the Proakis-C, the equal gain condition \cite{Bomfin} is not the only factor that determines the performance of a given waveform.
A detailed investigation on this topic is out of the scope of our work, however, a logical explanation is that the different \ac{ISI} patterns of both precoders at the receiver can be the cause for the performance discrepancy.


In terms of the complexity analysis shown in Tables \ref{tab:real_add} and \ref{tab:real_multi}, it is observed that for QPSK, there are approximately 25\% of additions and multiplications savings by using SDFT instead of the DFT. However, as the QAM order increases to 16-QAM and 64-QAM, the savings become negligible, because the constellation order $J$ plays a major role than the spreading parameter $Q$, as one can see in Table \ref{tab:complexity_analysis}.
The same effect happens when comparing SWH to SDFT. Basically, SWH can save approximately 25\% of multiplications for QPSK, but these savings also become negligible for higher $J$.

In conclusion, considering the performance gain and computational savings, the results of this analysis favor the SDFT precoder over DFT, especially for QPSK where the reduction in complexity is about 25\%. 
Lastly, the SDFT precoder also outperforms the SWH with the SILE-EPIC, however, a more general comparison is made in the subsequent subsection where the SWH is compared using the MAP receivers of Section \ref{sec:map_detectors_SWH}.



\subsection{Performance and Complexity trade-off Between (S)DFT and SWH Precoders}
\begin{figure*}[t!]
	\begin{minipage}{0.5\textwidth}
		\includegraphics[scale=0.85]{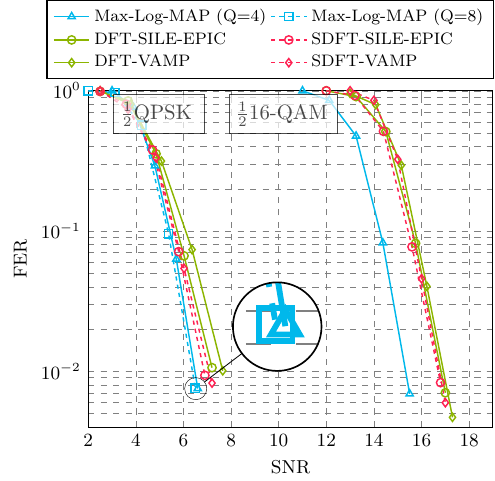}
	\end{minipage}
	\begin{minipage}{0.5\textwidth}
		\includegraphics[scale=0.5]{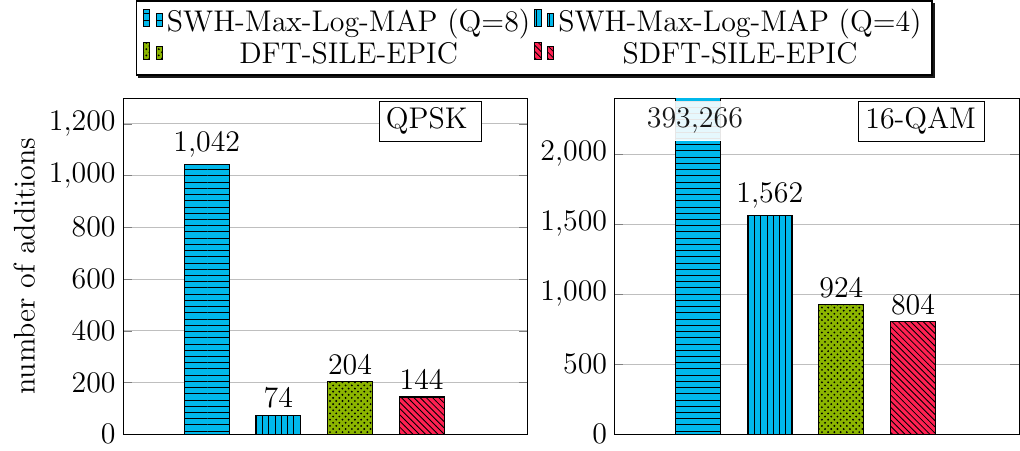}
\includegraphics[scale=0.5]{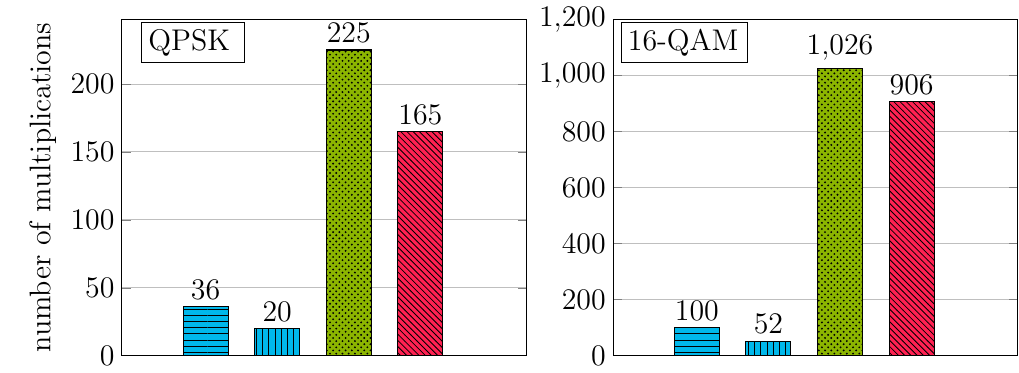}
	\end{minipage}

	\vspace{-0cm}
	\caption{\textcolor{black}{Performance and complexity trade-off between SWH-Max-Log-MAP and (S)DFT-SILE-EPIC.} \textcolor{black}{The FER results for the VAMP based equalizer \cite{VAMP} are shown for completeness.}}
	\label{fig:FER_ALL}
\end{figure*}

Although the SWH system performs worse than DFT system when both employ the SILE-EPIC, SWH has the advantage of being able to employ the MAP detector with reasonable implementation complexity.
Thus, it is worth comparing both systems with their best available receiver options, where the SWH with the Max-Log-MAP method is shown. These results are shown in Fig.~\ref{fig:FER_ALL}.

\textcolor{black}{For completeness, we include the results of the VAMP implementation \cite{VAMP} in this analysis. In particular, we implemented the LMMSE version described by Algorithm 3 in \cite{VAMP}, whose main difference from the SILE-EPIC is situated in the damping procedure of equation \eqref{eq:xa_va}, as noted in \cite{Sahin}. Specifically, the damping of VAMP is described in \cite[(26-27)]{VAMP}, which for the SILE-EPIC translates into combining $\mu^{\rm a}_n$ and $v^{\rm e}$ in \eqref{eq:mu_n} and \eqref{eq:v_e}, respectively, with their respective values of previous self-iterations. Lastly, since VAMP uses the same blocks as SILE-EPIC, its complexity is of the same order.
	In summary, the results depicted in Fig.~\ref{fig:FER_ALL} confirm the outcomes of \cite{Sahin}, where the SILE-EPIC shows superior performance than VAMP. In addition, we note that the performance gap is smaller for the SDFT than the DFT precoder, and the gap diminishes with higher QAM.}

For the QPSK and 16-QAM modulations, the SWH with MAP receivers have approximately 1dB and 2dB gains, respectively, over the DFT system for the FER of $10^{-2}$. This gap is slightly decreased when compared to the SDFT.
Regarding the complexity collected from Tables \ref{tab:real_add} and \ref{tab:real_multi}, we observe that for QPSK and $Q=4$, the Max-Log-MAP spends 74 additions and 20 multiplications, while the SDFT-SILE-EPIC system spends 144 additions and 165 multiplications.
Comparing both, we see that the Max-Log-MAP receiver spends 50\% on additions and 12\% on multiplications, which shows an overall complexity reduction.
However, for the configuration with $Q=8$, the Max-Log-MAP number of additions are multiplications are 1,042 and 36, respectively. While the number of multiplications remains low, the number of additions now is 7.24 times higher than the SDFT-SILE-EPIC, such that the complexity reduction is no more present.
Analyzing now the 16-QAM system with $Q=4$, the Max-Log-MAP method spends 1,562 additions and 52 multiplications, respectively, while the SDFT-SILE-EPIC spends 804 and 906. In absolute terms, the Max-Log-MAP spends 758 more additions and 662 fewer multiplications.
Given that multiplications are more costly computationally, we can conclude that in this case, the Max-Log-MAP provides lower complexity than the SDFT-SILE-EPIC for 16-QAM and $Q=4$.

In conclusion, the performance and complexity trade-off analysis favors the SWH precoder with Max-Log-MAP receiver when $Q=4$ is utilized for both QPSK and 16-QAM. For $Q=8$ with QPSK, the number of additions required by the Max-Log-MAP receiver surpasses the SILE-EPIC considerably, while the performance gain is marginal, meaning that setting $Q=4$ provides a good trade-off.
For all other configurations, namely, 16-QAM with $Q=8$ and 64-QAM, the Max-Log-MAP receiver complexity is very high in terms of additions, which limits its employment in those cases, which favors other implementations and/or precoders such as the SDFT-SILE-EPIC.

\section{Conclusion}\label{sec:conclusion}
In this work, we have provided two \textcolor{black}{feasible methods} for the SWH MAP detector, namely, Log-MAP and Max-Log-MAP.
While the Max-Log-MAP has a slight performance loss in relation to the Log-MAP, the number of additions and multiplications are decreased respectively by 1/2 and 1/3, which makes the Max-Log-MAP appealing in practice. 
Another aspect of this paper consisted in investigating the impact of SDFT, DFT, and SWH precoder employing the SILE-EPIC receiver.
The results have revealed that the SDFT system provides slightly better performance than the conventional DFT system with 25\% less complexity than the conventional DFT system for QPSK.
Lastly, the performance and complexity trade-offs between SWH-Max-Log-MAP and (S)DFT-SILE-EPIC have been investigated.
The results show that for QPSK and 16-QAM constellations, SWH-Max-Log has a better trade-off when setting the spreading parameter to $Q=4$.

For future work, we propose to investigate alternatives to decrease the complexity of the Max-Log-MAP detector for SWH even more, since its implementation is not feasible for $Q\geq 8$ with 16 order QAM or above, which limits its range of application.
Another interesting open point is the extension of the Max-Log-MAP for MIMO systems.
In particular, the detector provided in this work can be used to detect signal per transmit antenna with interference cancellation.
\textcolor{black}{Since the evaluation of this work is constrained to a highly frequency selective channel model, it is of interest to conduct this investigation under other types of channels, including time selectivity.}
\appendices
\section{Multiplications Reduction of MAP Detector}\label{appendix:MAP}
In the following, we show that the norm present in \eqref{eq:p_np} can be computed as
\begin{equation}\label{eq:norm}
	-\frac{1}{\sigma^2}\left \| {\mathbf{y}}^{\rm I}_{p}-|{\mathbf{\Lambda}}_p|\mathbf{W}_{Q}{\rm\mathbf{z}} \right \|^2 = \sum_{\substack{q=0, \\ i = \mathcal{S}(\ma{z},q)}}^{Q-1} C^{\rm I}_{p}(q,i).
\end{equation}
Let $\tilde{\mathbf{u}} = \tilde{\ma{W}}_Q\ma{z}$, where $\tilde{\ma{W}}_Q$ is the non normalized Walsh-Hadamard matrix of size $Q\times Q$, $\ma{z} \in \mathcal{R}^Q$ is the PAM vector of size $Q$ and
\begin{equation}
	\mathcal{R} = \left\{-\sqrt{J}+1+2i| i\in\left\{0,1,\cdots,\sqrt{J}-1\right\}\right\}
\end{equation}
is the PAM constellation set with cardinality $|\mathcal{R}|=\sqrt{J}$. 
It has been shown in \cite{BomfinMAP}, that all possible elements of $\tilde{\mathbf{u}}$ belong to the set
\begin{equation}\label{eq:Xcal}
	\mathcal{U} = \left\{-Q(\sqrt{J}\!-\!1) \!+ \!2i| i \!\in \left\{0,1,\cdots,Q(\sqrt{J}\!-\!1)\right\}\right\}
\end{equation}
that has cardinality $|\mathcal{U}| = Q(\sqrt{J}-1)+1$. The vector $\ma{s}$ with elements taken from $\mathcal{U}$ without repetition can be defined as 
\begin{equation}\label{eq:s}
	s_i = (-Q(\sqrt{J}\!-\!1) \!+ \!2i)/\sqrt{Q}
\end{equation}
for $i \in\left\{ 0,1,\cdots,Q(M\!-\!1)\right\}$. The division $\sqrt{Q}$ assumes a normalized WH transformation.

\textcolor{black}{Now, for a realization $\ma{z}$ and ${\mathbf{u}} = {\ma{W}}_Q\ma{z}$, we can define the quantity $C^{\rm I}_{p}(q,i)$ based on $s_i$ in \eqref{eq:s} as
\begin{equation}\label{eq:C_app}
	\begin{split}
		C^{\rm I}_{p}(q,i)  & = -\frac{1}{\sigma^2}\left({\mathbf{y}}^{\rm I}_{p}[q]-|{\mathbf{\Lambda}}_p[q,q]|{u}_q\right)^2
		\\ & =  -\frac{1}{\sigma^2}(\ma{y}^{\rm I}_{p}[q] - \ma{\Lambda}_p[q,q] s_i)^2,
	\end{split}
\end{equation}}
for $p \in \left\{0,1,\cdots, P-1\right\}$ and $q \in \left\{0,1,\cdots, Q-1\right\}$.

From \eqref{eq:C_app}, it is clear that \eqref{eq:norm} holds. In \eqref{eq:p_np} and \eqref{eq:norm}, the set $\mathcal{S}(\ma{z},q) = \{i | \textcolor{black}{u_q = s_i}\}$ returns the index $i$ which attains the condition \textcolor{black}{$u_q = s_i$} and can be interpreted as a database which stores the respective values for all $\ma{z} \in \mathcal{R}^Q$ and $q \in\left\{ 0,1,\cdots,Q-1\right\}$.


\section{Accuracy of Symbol Variance Approximation}\label{appendix:variance_approx}
The accuracy of $\bar{\gamma}^{\rm a}$ in \eqref{eq:gamma_a} and $\bar{\gamma}^{\rm a}_p$ in \eqref{eq:gamma_a_p} can be described in terms of their \acp{MSE}.
For the MSE of $\bar{\gamma}^{\rm a}$ we can write
\begin{equation}\label{eq:MSE}
	\begin{split}
\text{MSE} & = \frac{1}{N}\sum_{n=0}^{N-1} (\gamma^{\rm a}_n)^2-\left(\frac{1}{N}\sum_{n=0}^{N-1}\gamma^{\rm a}_n\right)^2 \\
		& = \frac{1}{N}\sum_{n=0}^{N-1} (\gamma^{\rm a}_n)^2-\left(\frac{1}{P}\sum_{p=0}^{P-1}\bar{\gamma}^{\rm a}_p\right)^2,
\end{split}
\end{equation}
where the second line is obtained by replacing $\frac{1}{N}\sum_{n=0}^{N-1}\gamma^{\rm a}_n$ by $\frac{1}{P}\sum_{p=0}^{P-1}\bar{\gamma}^{\rm a}_p$.
The average MSE for the system with sparsity $\bar{\gamma}^{\rm a}_p$ is written as
\begin{equation}\label{eq:MSE_s}
	\begin{split}
	\text{MSE}_{\rm s} & = \frac{1}{P}\sum_{p = 0}^{P-1}\left(\frac{1}{Q}\sum_{n\in I_p} (\gamma^{\rm a}_{n})^2-\left(\frac{1}{Q}\sum_{n\in I_p}\gamma^{\rm a}_n\right)^2\right), \\
	& = \frac{1}{N}\sum_{n=0}^{N-1} (\gamma^{\rm a}_n)^2 - \frac{1}{P}\sum_{p=0}^{P-1} \left(\bar{\gamma}^{\rm a}_p\right)^2.
\end{split}
\end{equation}
Our goal is to show that $\text{MSE}_{\rm s} \leq \text{MSE}$. From the above equations, this inequality implies $\left(\frac{1}{P}\sum_{p=0}^{P-1}\bar{\gamma}^{\rm a}_p\right)^2 \leq \frac{1}{P}\sum_{p=0}^{P-1} \left(\bar{\gamma}^{\rm a}_p\right)^2$, which holds due to Jensen's inequality since $f(x) = x^2$ is a convex function for $x \in \mathbb{R}_{\geq 0}$.

\bibliography{references}{}
\bibliographystyle{ieeetr}

\vfill

\end{document}